\documentclass{article}
\usepackage[preprint]{neurips_2023}

\usepackage[square,sort,comma,numbers]{natbib}
\usepackage[utf8]{inputenc} % allow utf-8 input
\usepackage[T1]{fontenc}    % use 8-bit T1 fonts
\usepackage{url}            % simple URL typesetting
\usepackage{booktabs}       % professional-quality tables
\usepackage{amsfonts}       % blackboard math symbols
\usepackage{nicefrac}       % compact symbols for 1/2, etc.
\usepackage{microtype}      % microtypography
\usepackage{xcolor}         % colors

\usepackage[hidelinks]{hyperref}
\usepackage{mathtools}
\usepackage{resizegather}
\usepackage{wrapfig}
\usepackage{tikz}
\usepackage{amsmath}
\usepackage{multirow}
\usepackage{color}
\usepackage{xcolor}
\usepackage{url}
\usepackage{caption}
\usepackage{subcaption}
\usepackage{float}
\usepackage{balance}
\usepackage{graphicx}
\usepackage{textcomp}
\usepackage{booktabs}
\usepackage{colortbl}
\usepackage[ruled,linesnumbered]{algorithm2e}  
\usepackage{enumerate}

\newcommand{\mypara}[1]{\smallskip\noindent\textbf{#1}}
\DeclareMathOperator*{\argmax}{arg\,max}

\newcommand{\softmax}{\mathrm{softmax}}

\title{Efficient Data-Free Model Stealing with Label Diversity}

\author{
\vspace{1mm}
Yiyong Liu \quad Rui Wen \quad Michael Backes \quad Yang Zhang \\
\vspace{1mm}
CISPA Helmholtz Center for Information Security \\
\texttt{\{yiyong.liu,rui.wen,director,zhang\}@cispa.de}
}

\begin{document}

\maketitle

\begin{abstract}
Machine learning as a Service (MLaaS) allows users to query the machine learning model in an API manner, which provides an opportunity for users to enjoy the benefits brought by the high-performance model trained on valuable data.
This interface boosts the proliferation of machine learning based applications, while on the other hand, it introduces the attack surface for model stealing attacks.
Existing model stealing attacks have relaxed their attack assumptions to the data-free setting, while keeping the effectiveness.
However, these methods are complex and consist of several components, which obscure the core on which the attack really depends.
In this paper, we revisit the model stealing problem from a diversity perspective and demonstrate that keeping the generated data samples more diverse across all the classes is the critical point for improving the attack performance. 
Based on this conjecture, we provide a simplified attack framework.
We empirically signify our conjecture by evaluating the effectiveness of our attack, and experimental results show that our approach is able to achieve comparable or even better performance compared with the state-of-the-art method.
Furthermore, benefiting from the absence of redundant components, our method demonstrates its advantages in attack efficiency and query budget.
\end{abstract}

\section{Introduction}
\label{section:introduction}
Machine Learning (ML) models have been deployed to perform a wide range of tasks with huge success.
However, targeting a well-generalized machine learning model is difficult as it requires tremendous amounts of time and money invested in both dataset collection and model training, which has become an obstacle on the road to the popularization of ML techniques.
To facilitate the use of ML techniques, companies make the trained model available as a service over the web (MLaaS), where the users can obtain the predictions with paid queries.
However, this poses a new threat to the confidentiality of the machine learning models, since the information comprised in the output enables the adversary to conduct malicious activities, e.g., performing model stealing attacks.

Model stealing attacks~\citep{OSF19} target to extract the functionality from the victim model and train a clone model locally.
The stolen model can even be leveraged for further attacks~\citep{SSSS17,ZJPWLS20,YLZZ19}.
To mount a model stealing attack, an adversary first queries the victim model to label the inputs, and then trains a clone model using these input-label pairs in a supervised manner. 
The quality of the queried samples has a significant impact on the performance of the clone model, and experimental results show that using random noise as input often leads to a model with unacceptable performance.
Previous attack methods~\citep{PMGJCS17,OSF19} utilize unlabeled samples from a similar distribution to query the model, which achieves nearly perfect clone model accuracy.
Unfortunately, it is hard to get access to such a dataset in practice, which limits the feasibility of the attack.

Recent attacks take a step further to relax this dataset assumption, i.e., they explore the possibility of stealing the model without the knowledge of the victim's training dataset distribution.
These attacks, which are known as data-free model stealing~\citep{TMWP21,KPQ21,SAB22}, mainly based on the idea of leveraging generative models~\citep{GPMXWOCB14,ACB17,ZGMO18,KLA19} to construct data samples that satisfy certain properties.
Specifically, some researchers~\citep{KPQ21,TMWP21} train a generator to synthesize difficult data samples by maximizing the disagreement between the victim model and clone model; while Sanyal et al.~\cite{SAB22} force the generator to fit the distribution of a proxy dataset. Though these attacks are effective, there indeed exists some problems.
First, the overall framework of such attacks is complicated, resulting in more computational cost. 
Second, the required query budget is much higher than the previous attacks with surrogate datasets, as it needs the prediction from the victim model for every generated image. 
Such drawbacks confine the attack efficiency and obscure the core property that makes the attacks work.

In this work, we refine the existing attack strategies and point out that diversity is the critical factor for model stealing.
Based on this conjecture, we provide a simplified attack framework from the angle of diversity, namely diversity-based data-free model stealing (DB-DFMS). 
Concretely, we take advantage of the generative models and force the generator to generate various images across all the classes, and our general hypothesis is that such images contain more information which can better represent the victim model's data distribution and thus enhance the attack performance. 
We conduct extensive experiments on three benchmark datasets, and the evaluation results demonstrate the effectiveness of our attack, which further confirms our conjecture.
Additionally, as our attack removes other redundant components, the attack exhibits economic advantages like requiring less query budget and being computationally friendly.
We further conduct our attack in more generalized settings, such as the attacker has no information about the clone model's architectures, which provides additional insights into understanding the success of our attack.

\section{Related Work}
\label{section:related_work}

\mypara{Model Stealing.}
Model stealing attack aims to extract the information from the victim model and constructs a local surrogate model. 
This attack was first proposed by Tramèr et al.~\cite{TZJRR16} where the adversary is assumed to have a surrogate dataset for stealing the model, While recent studies pay more attention to the most strict data-free setting, where no data is available for the adversary.
Under this scenario, Kariyappa et al.~\citep{KPQ21} propose MAZE, which uses a generative model to generate synthetic data samples for launching the attack. 
The generator is trained to maximize the disagreement between the victim model and the clone model, thus the gradients from the victim model are required.
They adopt zeroth order gradient estimation to approximate the gradients from the victim model as only black-box access is assumed here. 
A similar work, named DFME, is presented by Truong et al.~\citep{TMWP21}, while the key difference is to replace the loss function from Kullback-Leibler (KL) divergence to $\ell_1$ norm loss for training the student model.
Sanyal et al.~\citep{SAB22} go a step further to train a GAN with a synthetic dataset and utilize the gradients of the clone model as a proxy to the victim model's gradients, which we refer to as DFMS-SL.
In this paper, we focus on the same threat model as DFME and DFMS-SL, that is, we assume the attacker has no knowledge of the training dataset.

\mypara{Knowledge Distillation.}
Knowledge distillation~\citep{HVD15} is proposed to train a small student model efficiently with the knowledge from a large teacher model.
It uses softened output from the final layer of the teacher model as the label for training the student model.
In real scenarios, the training data of the teacher model is not available due to confidentiality, which motivates the concept of data-free knowledge distillation. 
Within this setting, the student has no information for the training data but access to the teacher model.
Nayak et al.~\citep{NMSRC19} exploit to obtain the prior information about the data distribution from the teacher model to craft data samples for training the student model. 
Most current works utilize the generative model to perform the distillation process~\citep{CCEL20,CWXYLSXXT19,MS19}.
They train the generative model with different loss objectives, targeting to synthesize data samples that are more aligned with the distribution of the teacher model.
However, all these distillation techniques require gradients from the teacher model, which is the major difference compared to data-free model stealing.

\section{Diversity-based data-free model stealing (DB-DFMS)}
\label{section:approach}
\subsection{Problem Statement}
In this paper, we focus on the problem of model stealing attack in the data-free setting.
In a nutshell, model stealing aims to train a local clone model $\mathcal{C}$ which is similar to the victim model $\mathcal{V}$.
The general attack workflow is as follows: the adversary has black-box access to the victim model, and they sample unlabeled data $x$ from a certain distribution.
For every unlabeled data $x$, the adversary queries the victim model $\mathcal{V}$ to obtain the prediction $\mathcal{V}(x)$. 
With the prediction $\mathcal{V}(x)$ and the corresponding input $x$, we can use ($x$, $\mathcal{V}(x)$)-pairs to form the surrogate dataset, which is used to train the clone model in a supervised way.

The distribution of the queried data has a significant influence on the performance of the clone model.
Orekondy et al.~\citep{OSF19} conduct model stealing attack by leveraging a surrogate dataset, but they fail to perform well if the surrogate dataset is not suitable to represent the distribution of the victim model.
Later, Roberts et al.~\citep{RPM19} even consider using random noise to launch the attack, however, results show that this attack cannot be generalized to sophisticated tasks like CIFAR-10.
In this paper, we consider the most challenging case where the adversary has no knowledge of the training dataset.

The problem studied in this paper has been explored in~\citep{KPQ21,TMWP21,SAB22}, their works propose data-free model stealing that can steal the model with high accuracy.
However, it is unclear what factor is the critical point that influences the quality of the clone model.
In this paper, we first reveal that diversity is the key to achieving good performance.
Based on this conjecture, we further propose a data-free model stealing attack that has comparable performance with a lower query budget and computational cost.

\begin{table*}[!t]
\centering
\setlength{\tabcolsep}{4.0pt}
\caption{The relationship between the entropy of the query dataset and the clone model accuracy for different model stealing attacks. 
The victim model is ResNet-34-8x trained on CIFAR-10 with a testing accuracy of 0.930, and the clone model is ResNet-18-8x.}
\scalebox{0.9}{
\begin{tabular}{lcccccc}
\toprule
\rowcolor{white}
Attack & \multicolumn{4}{c}{Surrogate Datasets} & \multicolumn{2}{c}{Data-free}\\
\cmidrule(l{5pt}r{5pt}){2-5}\cmidrule(l{5pt}r{0pt}){6-7} 
Scenarios & CelebA & SVHN & CIFAR-100 & CIFAR-10 & Random Noise & DB-DFMS (ours) \\
\midrule
Entropy (nats) & 1.05 & 1.10 & 2.16 & 2.30 & 1.20 & 1.95\\
Accuracy & 0.184 & 0.369 & 0.888 & 0.925 & 0.328 & 0.885\\
\bottomrule
\end{tabular}
}
\label{table:attack_for_surrogate_data_and_data_free}
\end{table*}

\subsection{Diversity is All You Need}
\label{section:intuition}
To have a better understanding of how the surrogate dataset influences the attack performance, we conduct model stealing attack on a ResNet-34-8x model trained on CIFAR-10 using different surrogate datasets.
As shown in \autoref{table:attack_for_surrogate_data_and_data_free}, we observe that CelebA has even worse performance compared with Random Noise, which rules out the hypothesis that more realistic images contribute more to the model performance.
There is still a widely accepted speculation that the more similar the data distribution is, the better the attack performance can be.
This speculation is partially proved in the table, as we can see that CIFAR-100 is the most similar to the training dataset CIFAR-10, which also has the best accuracy among the CelebA, SVHN, and CIFAR-100 dataset.

However, our experiments point out that this speculation is not entirely true, as the last column shows, our method could generate images (see \autoref{figure:generated_images}) that have low visual similarity to the training dataset, but leads to comparable performance as using CIFAR-100.
In other words, there is something more intrinsic behind it.

In this paper, we first point out that, the diversity of the query datasets (surrogate datasets
or synthetic datasets in data-free setting), defined as the entropy of the prediction probabilities from
the victim model, is the key point that influences the clone model performance, no matter in model
stealing attacks with surrogate datasets or in data-free setting.
Concretely, we calculate the diversity of each dataset in \autoref{table:attack_for_surrogate_data_and_data_free} and the results demonstrate a clear positive correlation.
Based on this observation, we design a diversity-based data-free model stealing attack in the following.

\subsection{Attack Pipeline}
\label{section:pipeline}
The attack workflow is shown in \autoref{figure:workflow}, and it can be divided into two entangled parts, i.e., the clone model training and the generator training.
In the following, we illustrate each part separately, then summarize them together.

\mypara{Clone Model Training.}
The training of the clone model follows the traditional one.
Concretely, the attack starts by taking a vector of random noise $z$ sampling from a normal distribution $\mathcal{N}(0,1)$ as input to the generator $\mathcal{G}$ and obtains a generated image $x$. 
Then the prediction $\mathcal{V}(x)$ can be acquired by querying the victim model $\mathcal{V}$, and the same to the prediction $\mathcal{C}(x)$ from the clone model $\mathcal{C}$. 
The clone model is trained to minimize the disagreement between $\mathcal{V}(x)$ and $\mathcal{C}(x)$. 
In this paper, we adopt $l_1$ distance to measure the agreement, since $l_1$ norm loss can prevent gradient vanishing, which has an advantage over KL divergence as shown in~\citep{TMWP21}. 
Formally, the loss for training the clone model is as follows:
\begin{equation}
    \label{equation:l1_loss}
    \mathcal{L}_{l_1} = \sum\limits_{i=1}^{K}\lvert \mathcal{V}_i(x) - \mathcal{C}_i(x)\rvert.
\end{equation}
where $K$ is the number of classes.
Note that $l_1$ norm loss requires the logits (i.e., the values before the $\softmax$ function), while we can only get the probability posteriors from the victim model. 
To address this issue, we follow the method proposed in~\citep{TMWP21} to approximate the logits from the probabilities, where we first calculate the logarithm of the probability vector and then subtract the log-probabilities with its mean value.

\SetKwComment{Comment}{/* }{ */}
\begin{minipage}{0.5\textwidth}
{\LinesNumberedHidden
\begin{algorithm*}[H]
\caption{DB-DFMS}
\label{algorithm:diversity}
\KwIn{Query budget $\mathcal{Q}$, generator iterations $n_{\mathcal{G}}$, clone iterations $n_{\mathcal{C}}$, learning rate $\eta$.}
\KwOut{Trained $\mathcal{C}$ and $\mathcal{G}$.}
\While{$\mathcal{Q} > 0$}{
    \For{$i = 1\cdots n_\mathcal{G}$}{
        $x = \mathcal{G}(z;\theta_{\mathcal{G}}), \quad \textrm{with} \quad z\sim \mathcal{N}(0,1)$\\
        $\alpha_k = \frac{1}{N}\sum\limits_{j=1}^{N}\softmax_k(\mathcal{C}(x_j))$\\
        $\mathcal{L}_{div} = \sum\limits_{k=1}^{K}\alpha_k \log \alpha_k$\\
        $\theta_{\mathcal{G}} = \theta_{\mathcal{G}} - \eta \nabla_{\theta_{\mathcal{G}}}\mathcal{L}_{div}$\\
    }
    \For{$i = 1\cdots n_{\mathcal{C}}$}{
        $x = \mathcal{G}(z;\theta_{\mathcal{G}}), \quad \textrm{with} \quad z\sim \mathcal{N}(0,1)$\\
        $\mathcal{L}_{l_1} = \frac{1}{N} \sum\limits_{j=1}^{N}\sum\limits_{k=1}^{K}\lvert \mathcal{V}_k(x_j) - \mathcal{C}_k(x_j)\rvert$\\
        $\theta_{\mathcal{C}} = \theta_{\mathcal{C}} - \eta \nabla_{\theta_{\mathcal{C}}}\mathcal{L}_{l_1}$
    }
    update query budget Q
}
\end{algorithm*}
}
\end{minipage}
\begin{minipage}{0.5\textwidth}
\centering
\captionsetup{type=figure}
\vspace{-3mm}
\includegraphics[width=0.9\columnwidth]{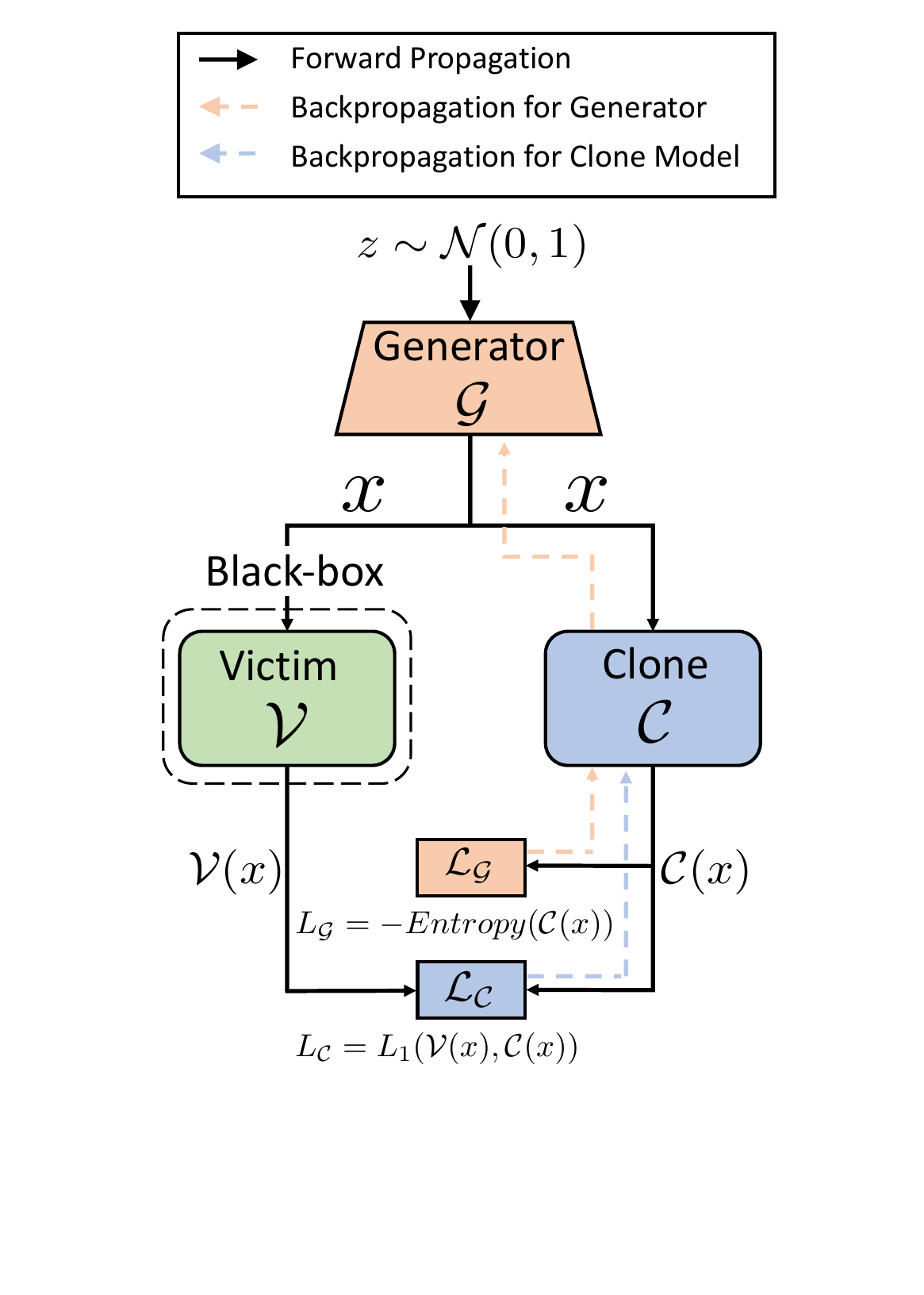}
\vspace{5pt}
\captionof{figure}{Workflow of DB-DFMS.}
\label{figure:workflow}
\end{minipage}

\mypara{Generator Training.}
Now we focus on the generator training part, which is vital as the generator determines the quality of generated samples.
As we discuss in the previous section, that diversity is the most important factor that influences the performance of model stealing, we want the generator to generate highly diverse images.
To achieve this goal, we use the negative entropy as the diversity loss to force the generation of more diverse images.
\begin{gather}
    \label{equation:diversity_loss}
    \mathcal{L}_{div}=\sum\limits_{i=1}^{K}\alpha_i\log\alpha_i, \quad \textrm{with} \quad \alpha_i = \frac{1}{N}\sum\limits_{j=1}^{N}\softmax_i(\mathcal{C}(x_j)).
\end{gather}
where $N$ is the batch size.
The diversity loss is calculated with the prediction from the clone model, as the victim model can only be accessed in black-box.

\mypara{Collaborative Training.}
As the diversity loss is calculated with the prediction from the clone model, 
therefore, the clone model is involved in the training of the generator.
Meanwhile, the training of the clone model also requires the contribution of the generator.
To solve this problem, we train the generator and clone model alternatively.
In order to better balance the training between the generator and clone model, for each iteration, the generator and clone model will be trained $n_{\mathcal{G}}$ and $n_{\mathcal{C}}$ times respectively.

\section{Evaluation}
\label{section:evaluation}
In this section, we empirically evaluate the effectiveness of our diversity-based data-free model stealing (DB-DFMS).
All experiments are performed on NVIDIA DGX A100 with Debian GNU/Linux 11.

\subsection{Experiment Setup}
\label{section:experiment_setup}
\mypara{Datasets.}
We conduct the experiments on three commonly used datasets: CIFAR-10~\citep{CIFAR}, SVHN, and CelebA~\citep{LLWT15}. 

\mypara{Model Training.}
We choose ResNet-34-8x~\citep{HZRS16} as the victim model architecture for all three tasks. 
We use ResNet-18-8x as the architecture of our clone model, and we explore the influence of model architectures in \autoref{section:clone_model_architecture}.
For the generator, we adopt the one used in~\citep{TMWP21}, which comprises three convolutional layers, together with linear up-sampling, batch normalization, and ReLU layers. 
To make the output lie in the range [-1,+1] (the predefined image domain), we add a hyperbolic tangent function to the last layer. 

\mypara{Attack Settings.}
We choose 2M query budget for SVHN and 20M for CIFAR-10 and CelebA to launch the attack.
And for training iterations of generator and clone model $n_\mathcal{G}$ and $n_\mathcal{C}$, we have tried different ratios, the general find is setting $n_\mathcal{C}$ a little bit higher than $n_\mathcal{G}$ can make the generator train smoothly and let the clone model see enough diverse data samples at the same time, thus we choose 1 and 5 as a suitable pair.

\mypara{Evaluation Metrics and Baselines.}
We choose accuracy and agreement to measure the quality of the clone model as they can directly demonstrate the similarity between the victim model and the clone model.
The training time of the clone model is also considered to reflect the attack efficiency. 
We compare our attack to two state-of-the-art methods, which we refer to as DFME~\citep{TMWP21} and DFMS-SL (with synthetic dataset)~\citep{SAB22}, and view the attack with random noise as a baseline. 

More details about the experiment setup can be found in \autoref{section:appendix_training}.

\begin{table*}[!t]
\centering
\caption{Performance of data-free model stealing against ResNet-34-8x trained on CIFAR-10, SVHN, and CelebA, ``Acc'' and ``Agr'' represent clone model accuracy and agreement between victim model and clone model respectively. The clone model is ResNet-18-8x.}
\label{table:main_attack_performance}
\setlength{\tabcolsep}{4.0pt}
\scalebox{0.9}{
\begin{tabular}{llcccccccc}
\toprule
\rowcolor{white}
Datasets & Victim & \multicolumn{2}{c}{Random Noise} & \multicolumn{2}{c}{DFME} & \multicolumn{2}{c}{DFMS-SL} & \multicolumn{2}{c}{DB-DFMS (Ours)} \\

\cmidrule(l{1pt}r{1pt}){3-4}\cmidrule(l{1pt}r{1pt}){5-6}\cmidrule(l{1pt}r{1pt}){7-8}\cmidrule(l{1pt}r{0pt}){9-10}
(budget) & accuracy & Acc & Agr & Acc & Agr & Acc & Agr & Acc & Agr\\
\midrule
CIFAR-10 (20M) & 0.930 & 0.328 & 0.314 & 0.869 & 0.893 & 0.896 & 0.926 & 0.885 & 0.921\\
SVHN (2M) & 0.962 & 0.808 & 0.814 & 0.952 & 0.971 & 0.955 & 0.977 & 0.955 & 0.975\\
CelebA (20M) & 0.769 & 0.706 & 0.759 & 0.750 & 0.865 & 0.743 & 0.813 & 0.746 & 0.853\\
\bottomrule
\end{tabular}
}
\end{table*}

\begin{table*}[!t]
\centering
\caption{Train time (s) of data-free model stealing against ResNet-34-8x trained on CIFAR-10, SVHN and CelebA. The Clone model is ResNet-18-8x.}
\label{table:training_time}
\setlength{\tabcolsep}{4.0pt}
\scalebox{0.9}{
\begin{tabular}{lcccc}
\toprule
\rowcolor{white}
Datasets (budget) & Random Noise & DFME & DFMS-SL & DB-DFMS (Ours) \\

\midrule
CIFAR-10 (20M) & 2749 & 3825 & $>$ 10000 & 3596\\
SVHN (2M) & 323 & 380 & $>$ 1000 & 374\\
CelebA (20M) & 8322 & 11759 & $>$ 30000 & 11147\\
\bottomrule
\end{tabular}
}
\end{table*}

\subsection{Effectiveness of DB-DFMS}
\label{section:results}
We conduct our experiments on three benchmark datasets and compare the attack performance of different data-free model stealing in \autoref{table:main_attack_performance}.
According to the clone model accuracy and agreement between the victim model and clone model, our attack can obtain a comparable attack result as other state-of-the-art methods on all three datasets.
These results demonstrate that with diversity loss only, the generated images are capable of contributing to a well-performed attack.

We further measure the computational cost consumed in the attack process with the training time, which is exhibited in \autoref{table:training_time}.
For Random Noise, it has the least training time as it doesn't need to train the generator, however, its attack performance is bad as shown in \autoref{table:main_attack_performance}. 
Among the rest methods, our attack saves the most computational time. 
Here the reason why DFMS-SL requires such high computation is it uses the proxy dataset to initialize the generator and clone model with hundreds of epochs.
However, when taking both the clone model performance and the training time into consideration, it is unnecessary to include such a proxy dataset in the attack process as it benefits so little and even has worse attack results in some datasets like CelebA.

To reach a deep comparison with DFMS-SL, we remove the initialization process of the generator and clone model, and directly train them alternatively with the proxy dataset. 
This simplified DFMS-SL requires a similar computational cost as our attack, however, the results downgrade accordingly. 
For example, the clone model accuracy on CIFAR-10 decreases from 0.896 to 0.849, which means the initialization process is crucial for DFMS-SL and the high computational cost is inevitable for it to obtain satisfactory attack performance.

\begin{table*}[!t]
\centering
\caption{Data-free model stealing with different clone model architectures against ResNet-34-8x trained on CIFAR-10, ``Acc'' and ``Agr'' represent clone model accuracy and agreement between victim model and clone model respectively.}
\label{table:different_clone_model}
\setlength{\tabcolsep}{4.0pt}
\scalebox{0.9}{
\begin{tabular}{llcccccccc}
\toprule
\rowcolor{white}
Architectures & Victim & \multicolumn{2}{c}{Random Noise} & \multicolumn{2}{c}{DFME} & \multicolumn{2}{c}{DFMS-SL} & \multicolumn{2}{c}{DB-DFMS (Ours)} \\

\cmidrule(l{1pt}r{1pt}){3-4}\cmidrule(l{1pt}r{1pt}){5-6}\cmidrule(l{1pt}r{1pt}){7-8}\cmidrule(l{1pt}r{0pt}){9-10}
(parameters) & accuracy & Acc & Agr & Acc & Agr & Acc & Agr & Acc & Agr\\
\midrule
MobileNetV2 (2.3M) & 0.930 & 0.264 & 0.252 & 0.819 & 0.815 & 0.870 & 0.894 & 0.853 & 0.881\\
DenseNet-121 (7.0M) & 0.930 & 0.309 & 0.311 & 0.863 & 0.876 & 0.885 & 0.898 & 0.875 & 0.892 \\
WideResNet-32 (7.4M) & 0.930 & 0.245 & 0.239 & 0.777 & 0.770 & 0.832 & 0.825 & 0.829 & 0.835\\
ResNet-18-8x (11.2M) & 0.930 & 0.328 & 0.314 & 0.869 & 0.893 & 0.896 & 0.926 & 0.885 & 0.921 \\
ResNet-34-8x (21.3M) & 0.930 & 0.308 & 0.297 & 0.883 & 0.900 & 0.905 & 0.929 & 0.891 & 0.922 \\
VGG-16BN (134.3M) & 0.930 & 0.191 & 0.194 & 0.699 & 0.677 & 0.793 & 0.770 & 0.789 & 0.772\\
\bottomrule
\end{tabular}
}
\end{table*}
\subsection{Influence of the Clone Model Architecture}
\label{section:clone_model_architecture}
As investigated in the previous work about knowledge distillation~\citep{CH19,MS19}, a smaller student model is sufficient to distill the knowledge from the teacher model, as long as it has enough capability. 
Thus we choose Resnet-18-8x as the clone model though the victim model is ResNet-34-8x. 
However, we are still interested in whether the clone model performance will be improved with higher capability.
Except Resnet-18-8x and ResNet-34-8x, we test other 4 commonly used model architectures, including MobileNetV2~\citep{SHZZC18}, DenseNet-121~\citep{HLMW17}, WideResNet-32~\citep{ZK16} and VGG-16BN~\citep{SZ15}.

In general. the attack performance is increased as the clone model has more number of parameters (See \autoref{table:different_clone_model}). 
However, there are two exceptions. 
The performance of DenseNet-121 is pretty well though its capability is comparatively low, which can be explained by its specific design.
That is, any two layers in the model are connected together to strengthen feature propagation. 
The other case is VGG-16BN, though it has hundreds of millions of parameters, it cannot obtain a satisfying attack result, which may be due to the more obvious differences in architectures to other models.
Compared to DenseNet-121, WideResNet-32 is more similar to the victim model ResNet-34-8x, however, the gap between the performance of these two models rules out that more similar network achieves better attack performance.

\subsection{Influence of the Query Budget}
\label{section:query_budget}
In previous experiments, we set the query budget to 20M for CIFAR-10, which is a common setting in previous works~\citep{KPQ21,TMWP21}.
In this section, we aim to explore how the query budget influences the attack performance.
A thorough understanding of the influence of query budget is beneficial as a lower query budget requires less computational power and lower money payment to the MLaaS platform.

The results are given in \autoref{figure:query_budget}.
As we can see, all the attack performance has a positive correlation to the query budget, both holding for the clone model accuracy and the agreement between the victim model and the clone model, while the margin increase above 10M is not that large.
However, our attack still consistently performs well and shows its advantage over other methods in some cases of query budget.
Especially when there is some query limitation like under 10M, our attack will be more efficient.

%-------------------------------------------------------------------------------
\begin{minipage}{0.45\textwidth}
    \centering
    \captionsetup{type=figure}
    \subfloat[Accuracy]{\includegraphics[width=0.5\linewidth]{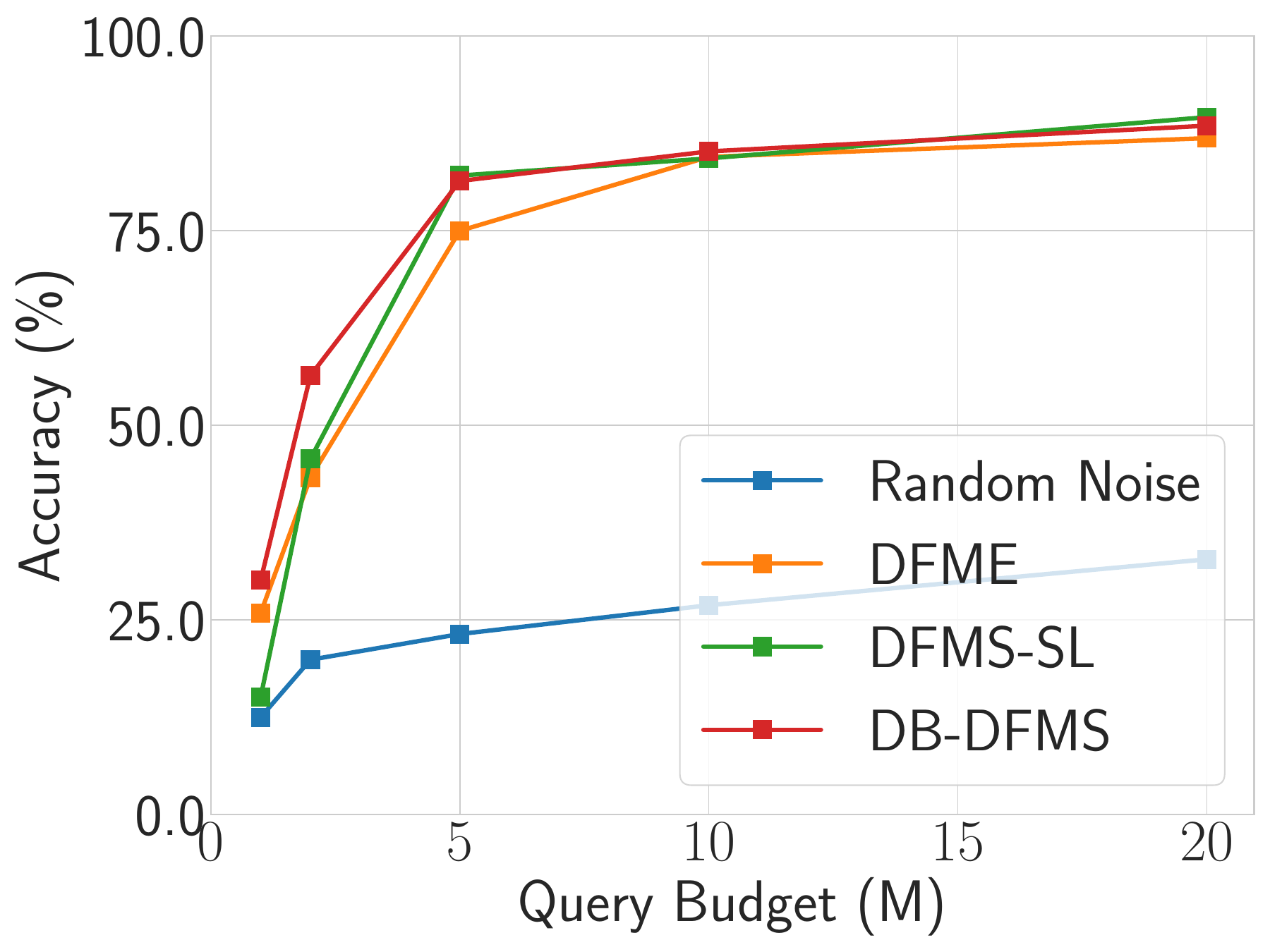}}
    \subfloat[Agreement]{\includegraphics[width=0.5\linewidth]{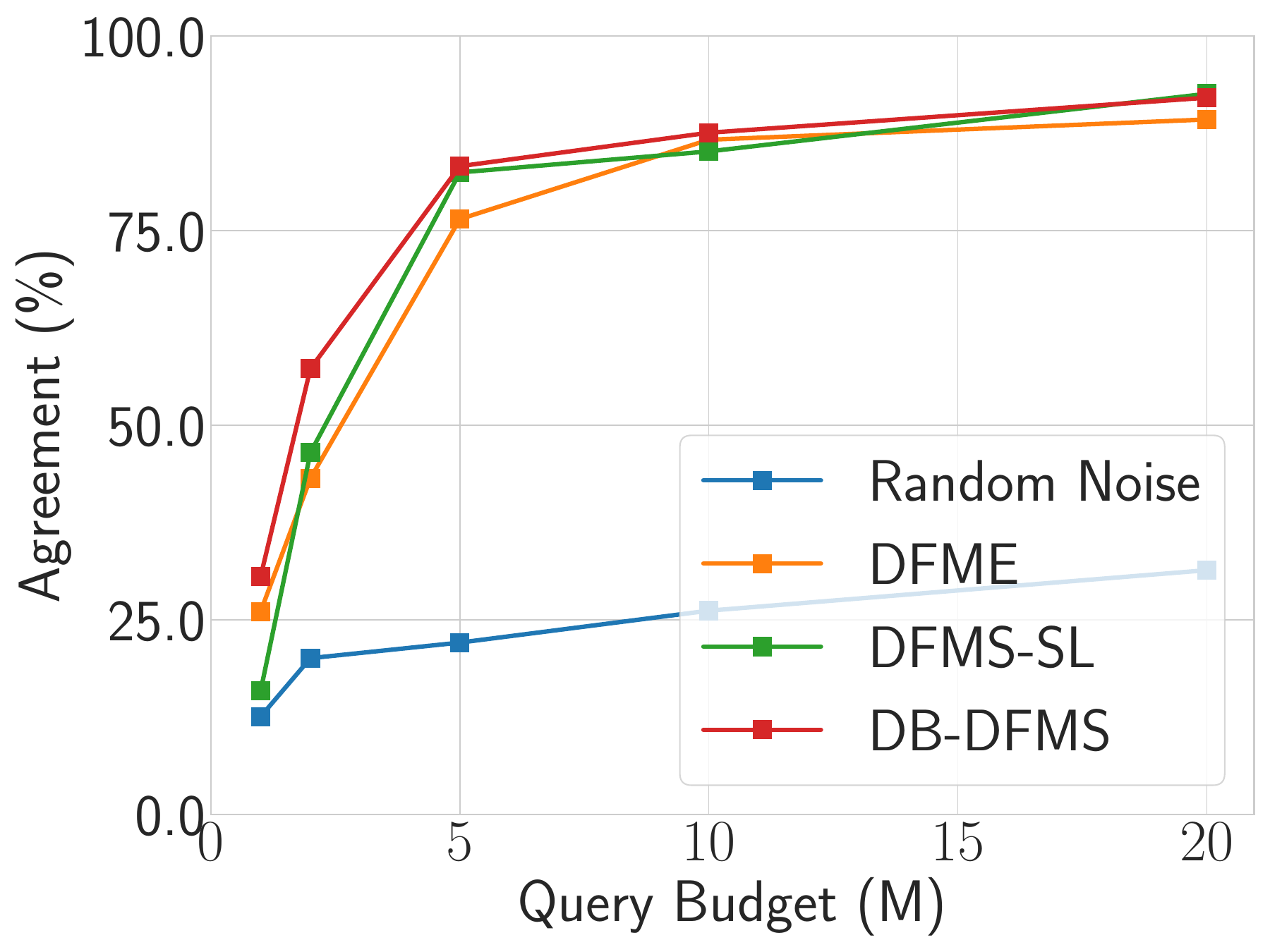}}
    \caption{The attack performance of DB-DFMS against ResNet-34-8x trained on CIFAR-10 with different query budget. The clone model is ResNet-18-8x.}
    \label{figure:query_budget}
\end{minipage}
%-------------------------------------------------------------------------------
\hfill
%-------------------------------------------------------------------------------
\begin{minipage}{0.45\textwidth}
    \centering
    \captionsetup{type=figure}
    \subfloat[Accuracy]{\includegraphics[width=0.5\linewidth]{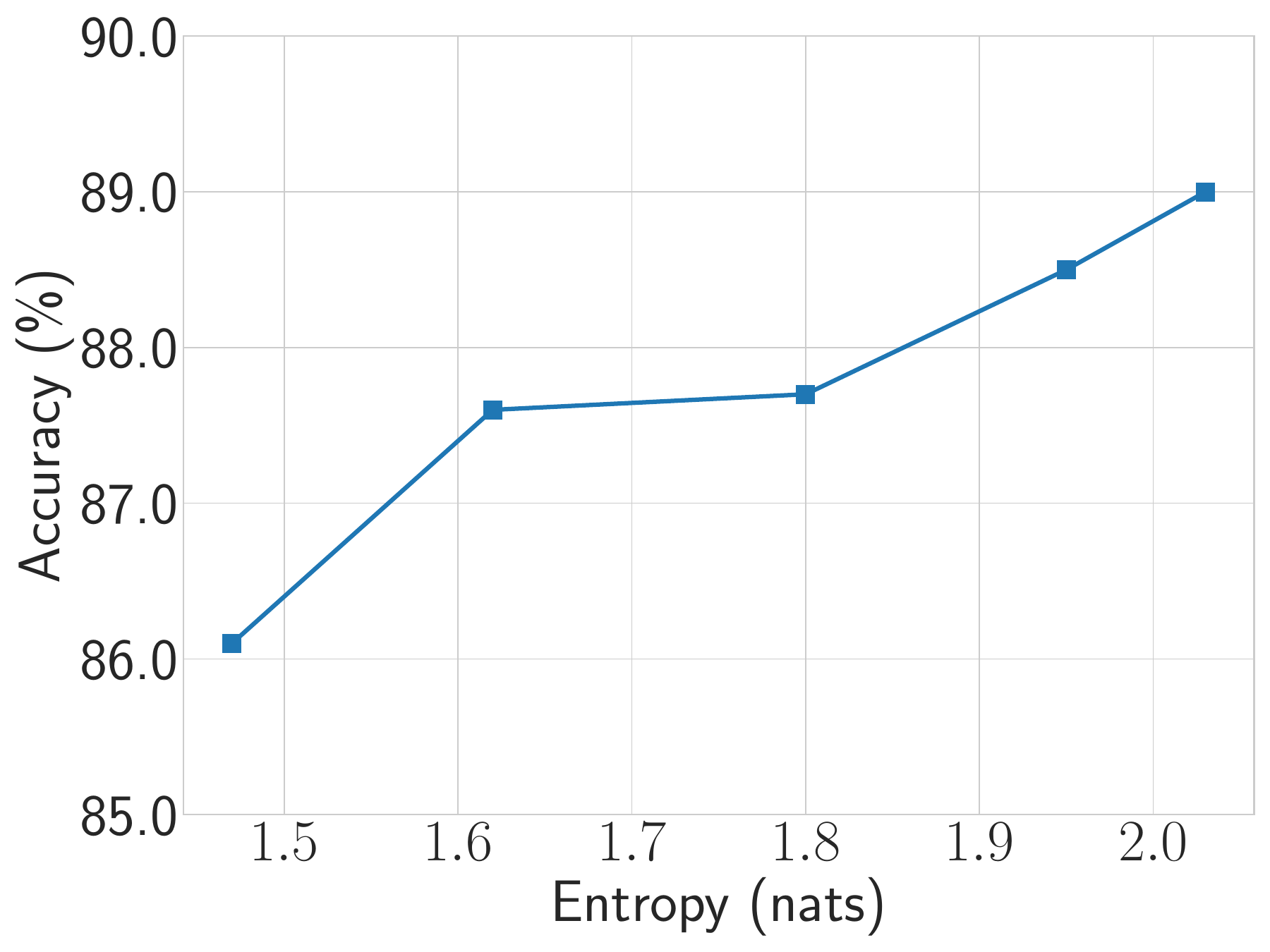}}
    \subfloat[Agreement]{\includegraphics[width=0.5\linewidth]{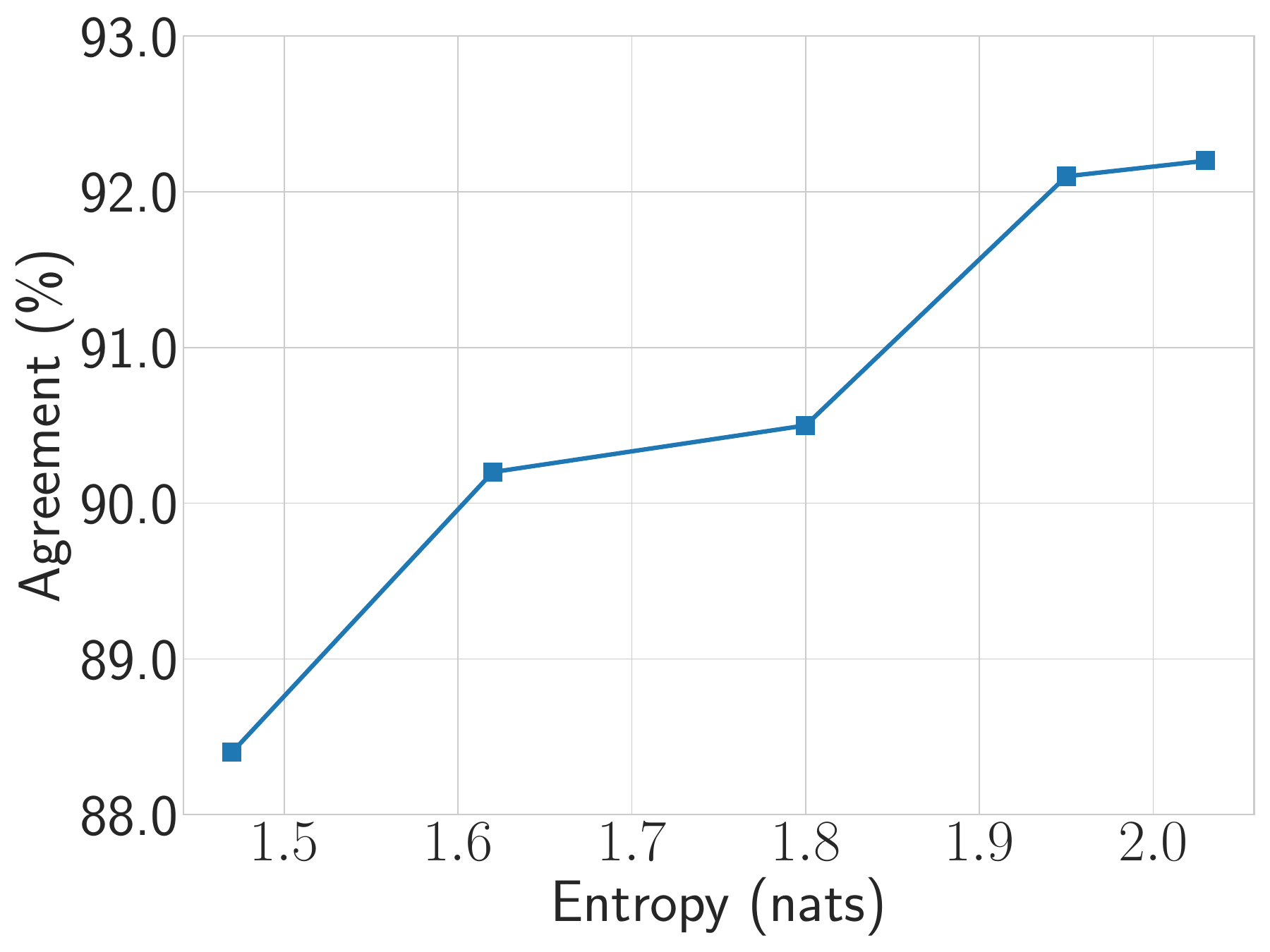}}
    \caption{Effect of generator architecture for entropy and attack performance. The victim model is ResNet-34 trained on CIFAR-10 and the clone model is ResNet-18-8x.}
    \label{figure:generator_entropy}
\end{minipage}
%-------------------------------------------------------------------------------

\subsection{Influence of the Generator Architecture}
\label{section:generator_architecture}

Apart from the clone model, the capacity of the generator should also have effect on the attack performance by influencing the quality of the generated images. 
Thus we vary the architecture of the generator by changing the number of convolutional blocks and the size of dimensions, and demonstrate the relationship between the attack results and the corresponding entropy of the generated data samples.

First, we find that the existence of convolutional layers is crucial for generating images with high diversity.
That is, if we remove all the convolutional blocks in the generator and leave only the linear layers, the attack performance is on par with directly using the random noise as shown in \autoref{table:attack_for_surrogate_data_and_data_free}, which means the linear transformation has nearly no contribution to enriching the diversity of the generated images.
Other results for generator with convolutional blocks are reported in \autoref{figure:generator_entropy}. 
It is clear that there is a close correlation between the attack performance and the entropy of the generated images, which further proves the importance of using images with high diversity in model stealing attacks.

\subsection{Influence of the Diversity Loss}
\label{section:types_of_diversity_loss}
The core idea of our attack is to train a generator by simply using a diversity loss, and the definition of diversity could be interpreted in different ways.
In this section, we form two different types of diversity loss and evaluate their impact on the attack performance.

\mypara{Sample Level.}
As shown in \autoref{equation:diversity_loss}, after the $\softmax$ function, the original loss first calculates the mean value over the batch of data samples, and then gets the negative entropy. 
Our first thought is to change the order of these two calculations and compute the entropy over each data sample first then average the entropy on all samples, thus we call it ``Sample Level'' diversity loss:
\begin{gather}
    \label{equation:sample_level_diversity_loss}
    \mathcal{L}_{sl\_div} = \frac{1}{N}\sum\limits_{i=1}^{N}\sum\limits_{j=1}^{K}\alpha_{ij}\log{\alpha_{ij}}, \quad \text{with} \quad \alpha_{ij} = \softmax_j(\mathcal{C}(x_i)).
\end{gather}

\mypara{Label Level.}
As the goal of diversity loss is to help the generation of more diverse data samples across all the classes, thus we try to calculate the diversity in a more direct way, i.e., by utilizing the hard labels:
\begin{gather}
    \label{equation:hard_label_diversity_loss}
    \mathcal{L}_{hl\_{div}} = \sum\limits_{i}^{K}\alpha_{i}\log\alpha_{i}, \quad \text{with} \quad \alpha_{i} = \frac{1}{N}\sum\limits_{j=1}^{N}\mathcal{F}_{one\_hot}(\argmax(\softmax(\mathcal{C}(x_{j}))), K)_{i}.
\end{gather}
where the $\argmax$ is to obtain the class index with the highest prediction probability, and $\mathcal{F}_{one\_hot}$ is a function to form a one-hot vector according to the class index and the total number of classes.
The difference between this loss and the original one is this loss does not take sample-wise diversity into consideration, but only tries to generate more samples belonging to different classes.

%-------------------------------------------------------------------------------
\begin{minipage}{0.45\textwidth}
    \centering
    \captionsetup{type=figure}
    \subfloat[Accuracy]{\includegraphics[width=0.5\linewidth]{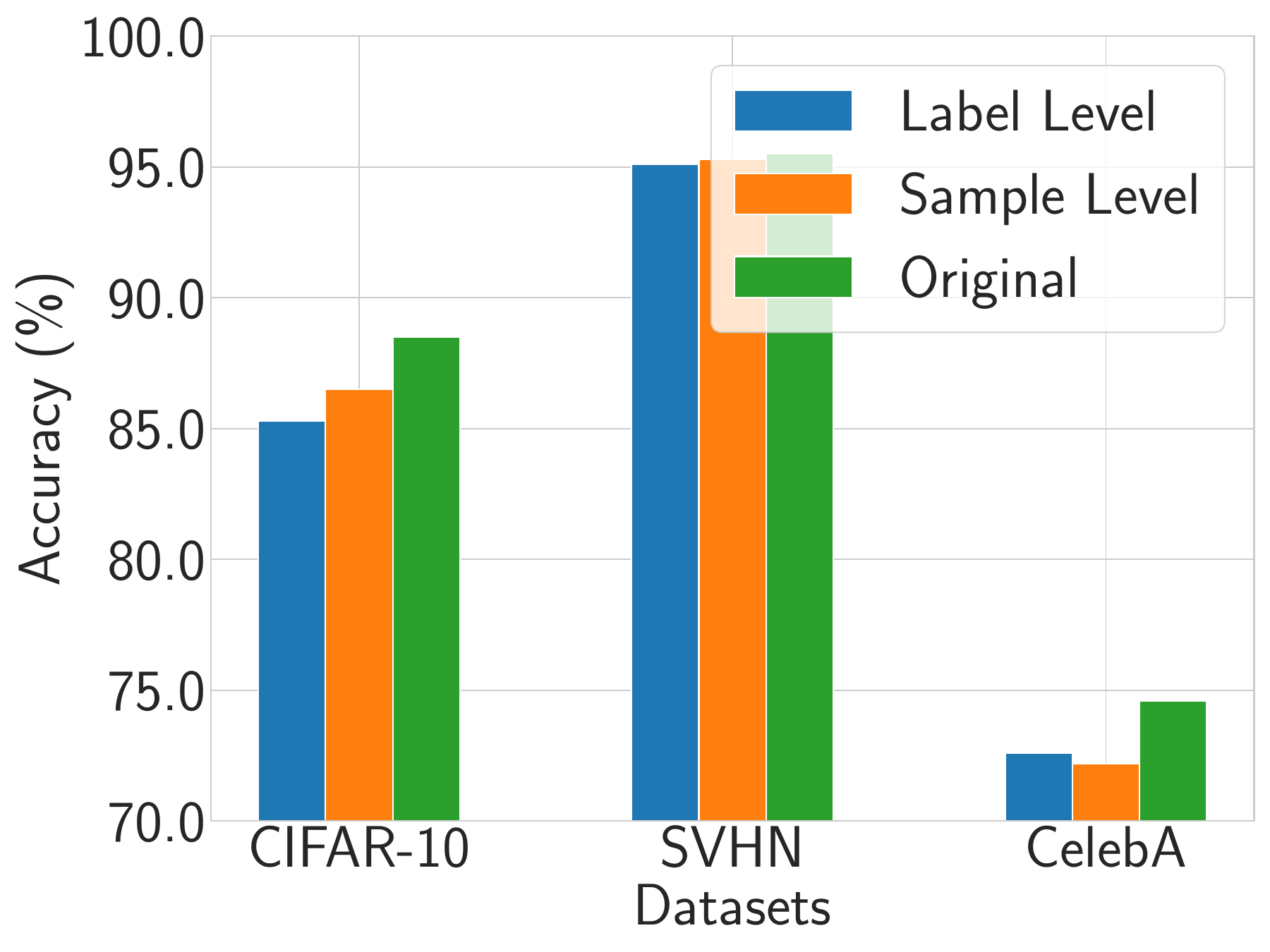}}
    \subfloat[Agreement]{\includegraphics[width=0.5\linewidth]{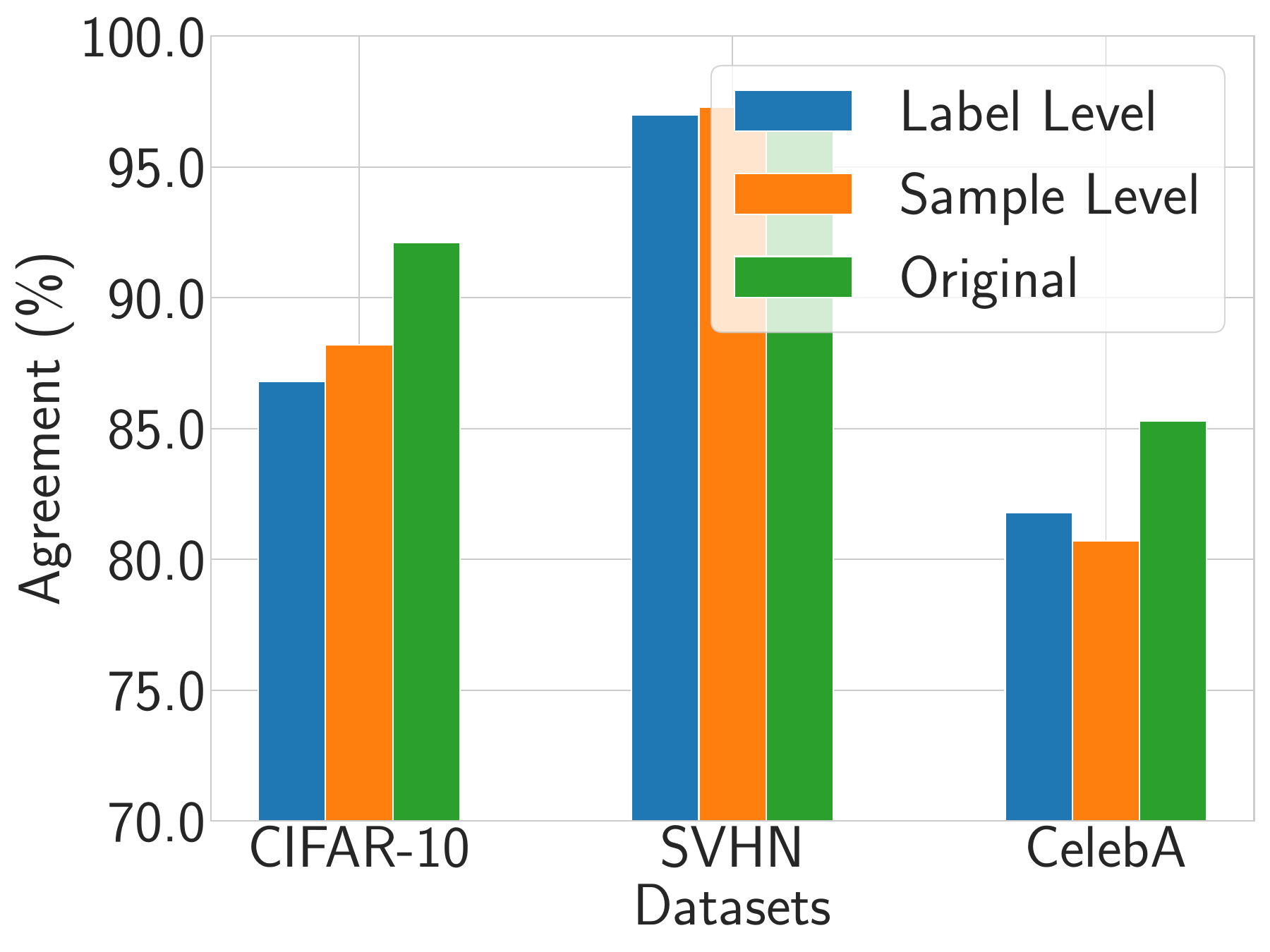}}
    \caption{The attack performance of DB-DFMS with different diversity loss against ResNet-34-8x trained on CIFAR-10. The clone model is ResNet-18-8x.}
    \label{figure:different_div}
\end{minipage}
%-------------------------------------------------------------------------------
\hfill
\begin{minipage}{0.45\textwidth}
\vspace{-4mm}
\centering
\captionsetup{type=table}
\setlength{\tabcolsep}{5.0pt}
\resizebox{0.95\textwidth}{!}
{
\begin{tabular}{l|ccc}
\toprule
Methods & Time (s) & Acc & Agr\\
\midrule
Random Noise & 2584 & 0.404 & 0.410 \\
DFME & 3629 & 0.776 & 0.788 \\
DFMS-SL & > 10000 & 0.766 & 0.777 \\
DB-DFMS (Ours) & 3350 & 0.783 & 0.796\\
\bottomrule
\end{tabular}
}
\vspace{9mm}
\caption{Attack performance of different methods against ResNet-34-8x trained on the unbalanced version of CIFAR-10.}
\label{table:acc_unbalanced}
\end{minipage}

We put the results in \autoref{figure:different_div}, results show that all three losses achieve good attack performance.
And we find that the best is still the one with the original diversity loss for all three datasets and the two evaluation metrics. 
Its advantage is derived from more information it utilizes as it considers all data samples in a mini-batch together and uses posteriors from the clone model.
We leave the finding of more elegant diversity loss as future work.

\subsection{Performance on Unbalanced Data}
\label{section:unbalanced_data}
Two datasets CIFAR-10 and CelebA used in our main experiments are balanced where each class has the same number of data samples. 
The only unbalanced data SVHN is simple as all methods have quite good performance.
Here we manually create an unbalanced dataset from CIFAR-10.
Specifically, for class from ``0'' to ``9'', we increase the number of data samples from $3200$ to $5000$, and all the data are selected randomly.
As shown in~\autoref{table:acc_unbalanced}, our method outperforms the two state-of-the-art attacks for both model accuracy and agreement with less training time. 
It indicates that though our attack aims to generate images with larger label diversity, it still works for the unbalanced data, which means our attack can be applied without prior knowledge of whether the classifier is trained on balanced data or not.

\begin{minipage}{0.45\textwidth}
\centering
\captionsetup{type=figure}
\includegraphics[width=0.7\textwidth]{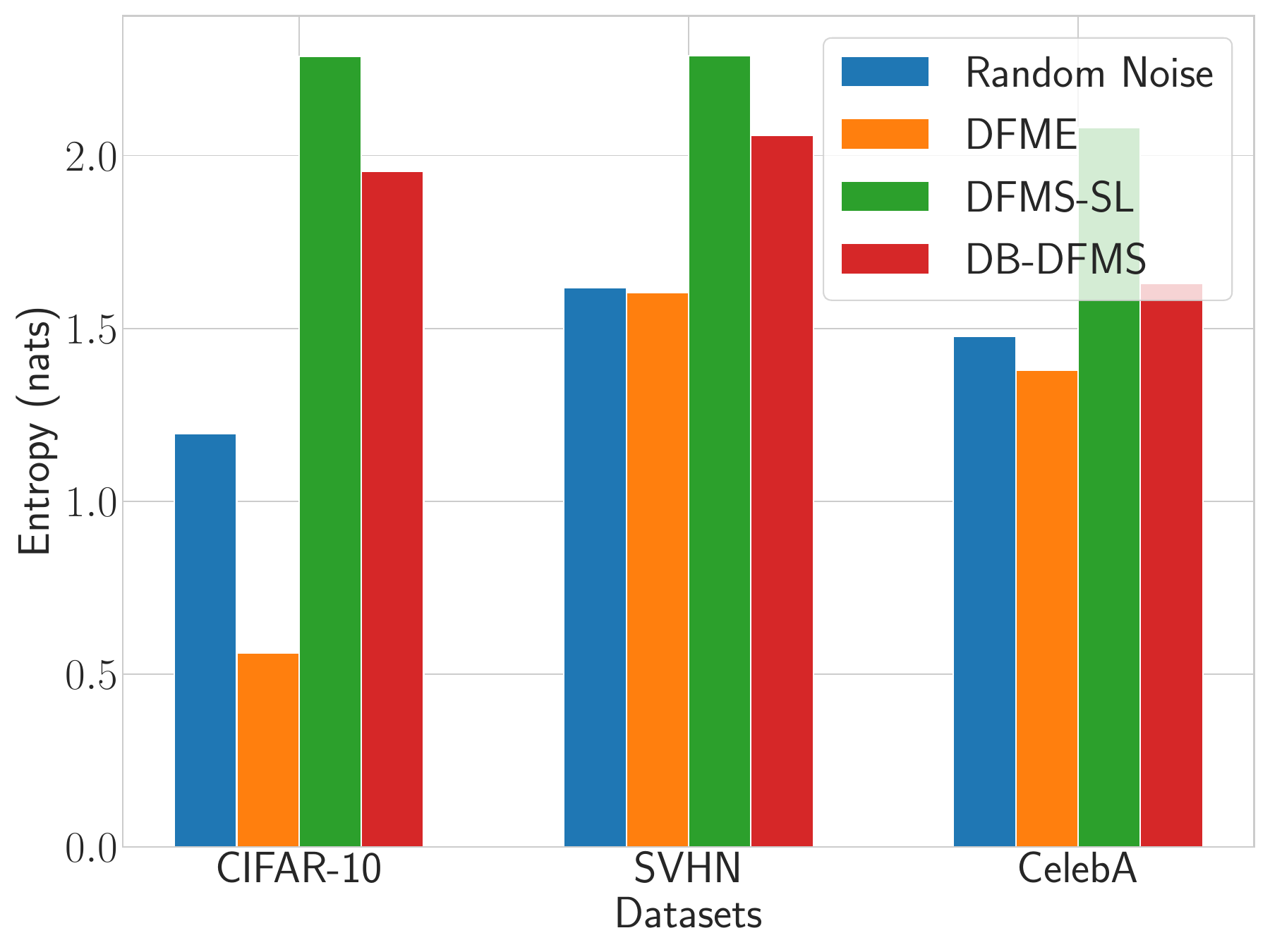}
\vspace{6mm}
\captionof{figure}{Entropy of generated data samples according to the prediction from victim model. The victim model is ResNet-34-8x trained on CIFAR-10 and clone model is ResNet-18-8x.}
\label{figure:entropy}
\end{minipage}
\hfill
\begin{minipage}{0.45\textwidth}
\centering
\captionsetup{type=figure}
\captionsetup[subfigure]{labelformat=empty}
\hspace{-1mm}
\subfloat{\includegraphics[width=0.11\columnwidth]{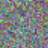}
\includegraphics[width=0.11\columnwidth]{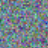}
\includegraphics[width=0.11\columnwidth]{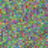}
\includegraphics[width=0.11\columnwidth]{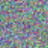}}
\hspace{1mm}
\subfloat{\includegraphics[width=0.11\columnwidth]{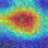}
\includegraphics[width=0.11\columnwidth]{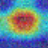}
\includegraphics[width=0.11\columnwidth]{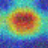}
\includegraphics[width=0.11\columnwidth]{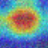}}

\hspace{-1mm}
\subfloat{\includegraphics[width=0.11\columnwidth]{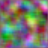}
\includegraphics[width=0.11\columnwidth]{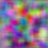}
\includegraphics[width=0.11\columnwidth]{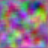}
\includegraphics[width=0.11\columnwidth]{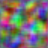}}
\hspace{1mm}
\subfloat{\includegraphics[width=0.11\columnwidth]{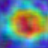}
\includegraphics[width=0.11\columnwidth]{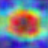}
\includegraphics[width=0.11\columnwidth]{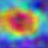}
\includegraphics[width=0.11\columnwidth]{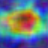}}

\hspace{-1mm}
\subfloat{\includegraphics[width=0.11\columnwidth]{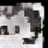}
\includegraphics[width=0.11\columnwidth]{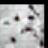}
\includegraphics[width=0.11\columnwidth]{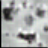}
\includegraphics[width=0.11\columnwidth]{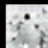}}
\hspace{1mm}
\subfloat{\includegraphics[width=0.11\columnwidth]{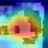}
\includegraphics[width=0.11\columnwidth]{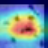}
\includegraphics[width=0.11\columnwidth]{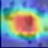}
\includegraphics[width=0.11\columnwidth]{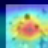}}

\hspace{-1mm}
\subfloat[Generated Images]{\includegraphics[width=0.11\columnwidth]{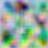}
\includegraphics[width=0.11\columnwidth]{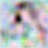}
\includegraphics[width=0.11\columnwidth]{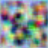}
\includegraphics[width=0.11\columnwidth]{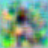}}
\hspace{1mm}
\subfloat[Grad Cam]{\includegraphics[width=0.11\columnwidth]{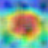}
\includegraphics[width=0.11\columnwidth]{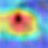}
\includegraphics[width=0.11\columnwidth]{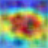}
\includegraphics[width=0.11\columnwidth]{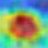}}

\vspace{16pt}
\caption{Generated data Samples and visualization of Grad-CAM from Random Noise, DFME, DFMS-SL and DB-DFMS (from top to bottom) for models trained on CIFAR-10.}
\label{figure:generated_images}
\end{minipage}

\section{Exploration}
\label{section:exploration}
Our simplified attack has comparable clone model performance and needs less training time, here we provide deep insights to show why the diversity of the generated data samples is the key point for enhancing the attack.

\subsection{Entropy of Generated Dataset}
\label{section:entropy}
As shown in \autoref{table:attack_for_surrogate_data_and_data_free}, the attack performance has a positive correlation to the diversity of the generated images, and here we further prove this finding. \autoref{figure:entropy} reports that DFMS-SL and our DB-DFMS have impressive clone model performance as both of them enable the generation of high entropy data samples.
We also admit that the diversity of the query datasets is not the only factor that influences the attack results.
We can find that DFME also performs well though the entropy of its generated dataset is comparatively low, which means the output of more difficult query samples from the victim model can lead to clone model with high performance as well. 

\subsection{Visualization of Generated Dataset}
\label{section:visualization}
We then visualize the generated data samples and the corresponding Grad-cam in~\autoref{figure:generated_images}.
As for generated images, ``Random Noise'' generates each pixel randomly, which means the neighboring pixels do not have correlation either, thus resulting in a grainy image.
While the rest images generated from other methods are comparably more smoothed.
To better understand how the machine learning model interprets these generated images, we adopt ``Grad CAM''~\citep{SCDVPB17} to localize the regions that contribute the most to the prediction.
As demonstrated in \autoref{figure:generated_images}, the ``Grad CAM'' visualizations follow almost the same pattern for different ``Random Noise'' generated images. 
The images from ``DFME'' show a bit difference compared to those from ``Random Noise''.
The most obvious changes can be found in the images generated by ``DFMS-SL'' and ``DB-DFMS''.
The central region related to the predictions are more diverse, which indicates different features can be learned from different images and thus more information can be utilized to boost the clone model performance.

\begin{figure*}[t]
\centering
\begin{subfigure}{0.245\columnwidth}
\includegraphics[width=\columnwidth]{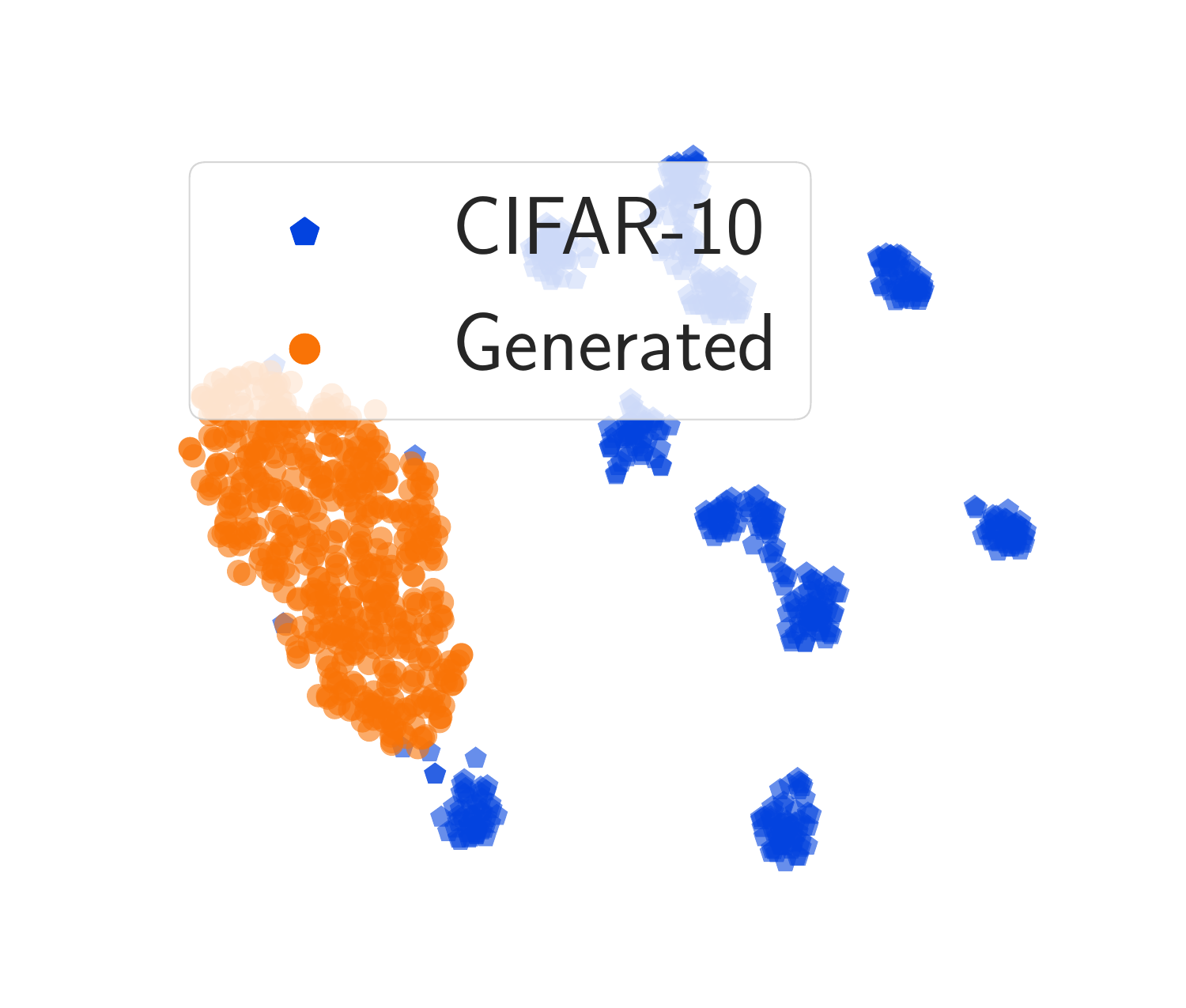}
\caption{Random Noise}
\label{figure:noise_tsne_embeddings}
\end{subfigure}
\begin{subfigure}{0.245\columnwidth}
\includegraphics[width=\columnwidth]{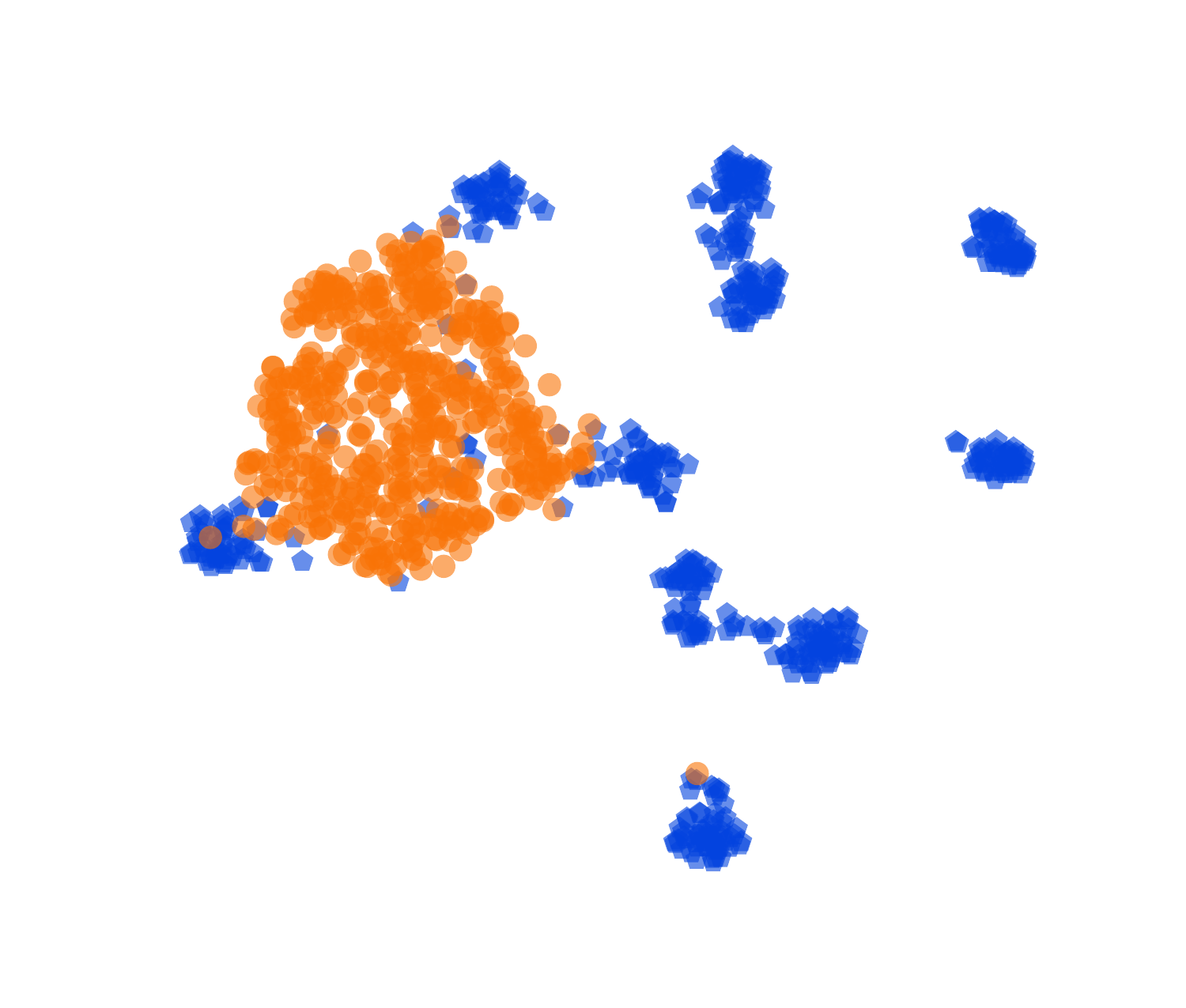}
\caption{DFME}
\label{figure:dfme_tsne_embeddings}
\end{subfigure}
\begin{subfigure}{0.245\columnwidth}
\includegraphics[width=\columnwidth]{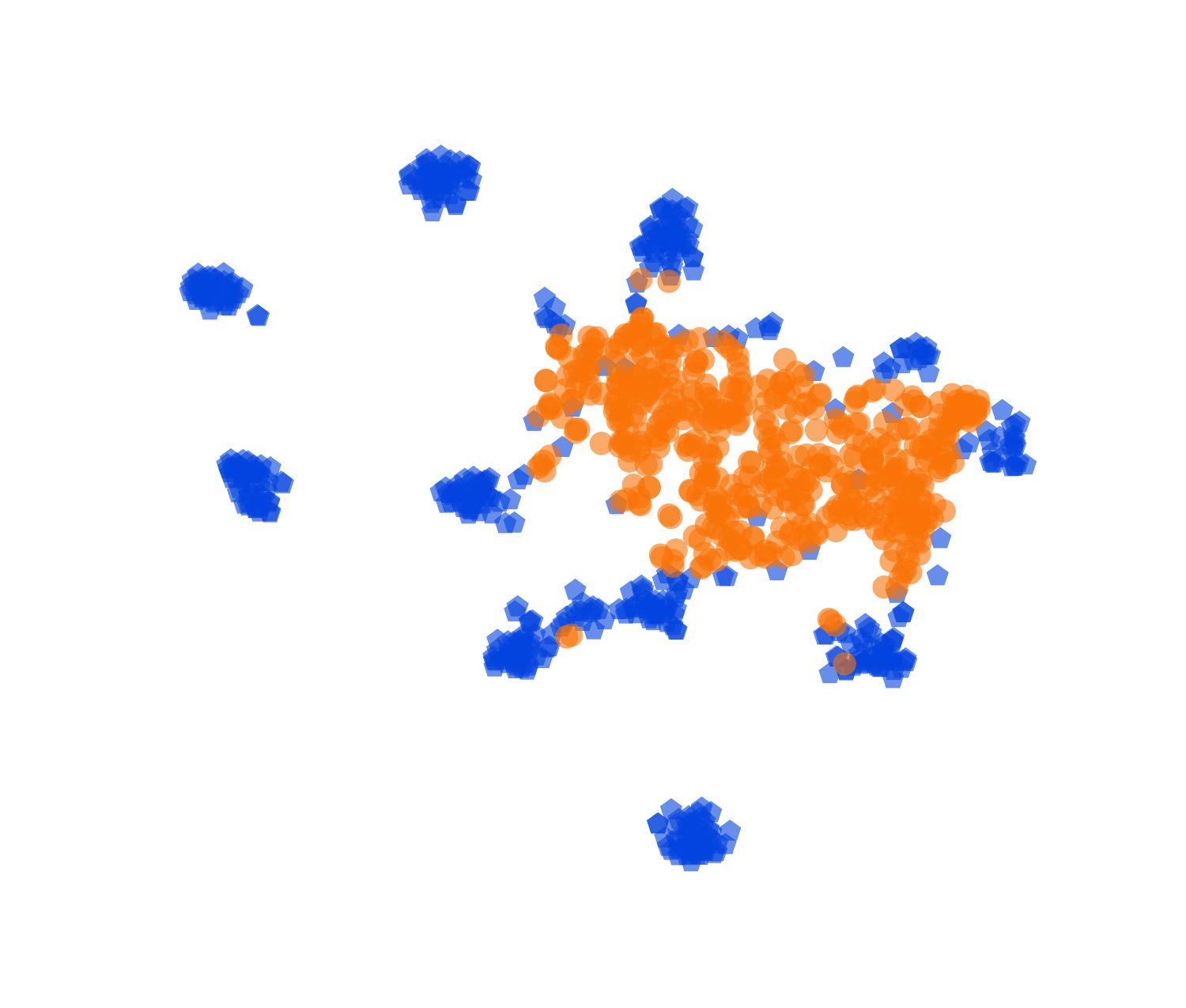}
\caption{DFMS-SL}
\label{figure:dfms_tsne_embeddings}
\end{subfigure}
\begin{subfigure}{0.245\columnwidth}
\includegraphics[width=\columnwidth]{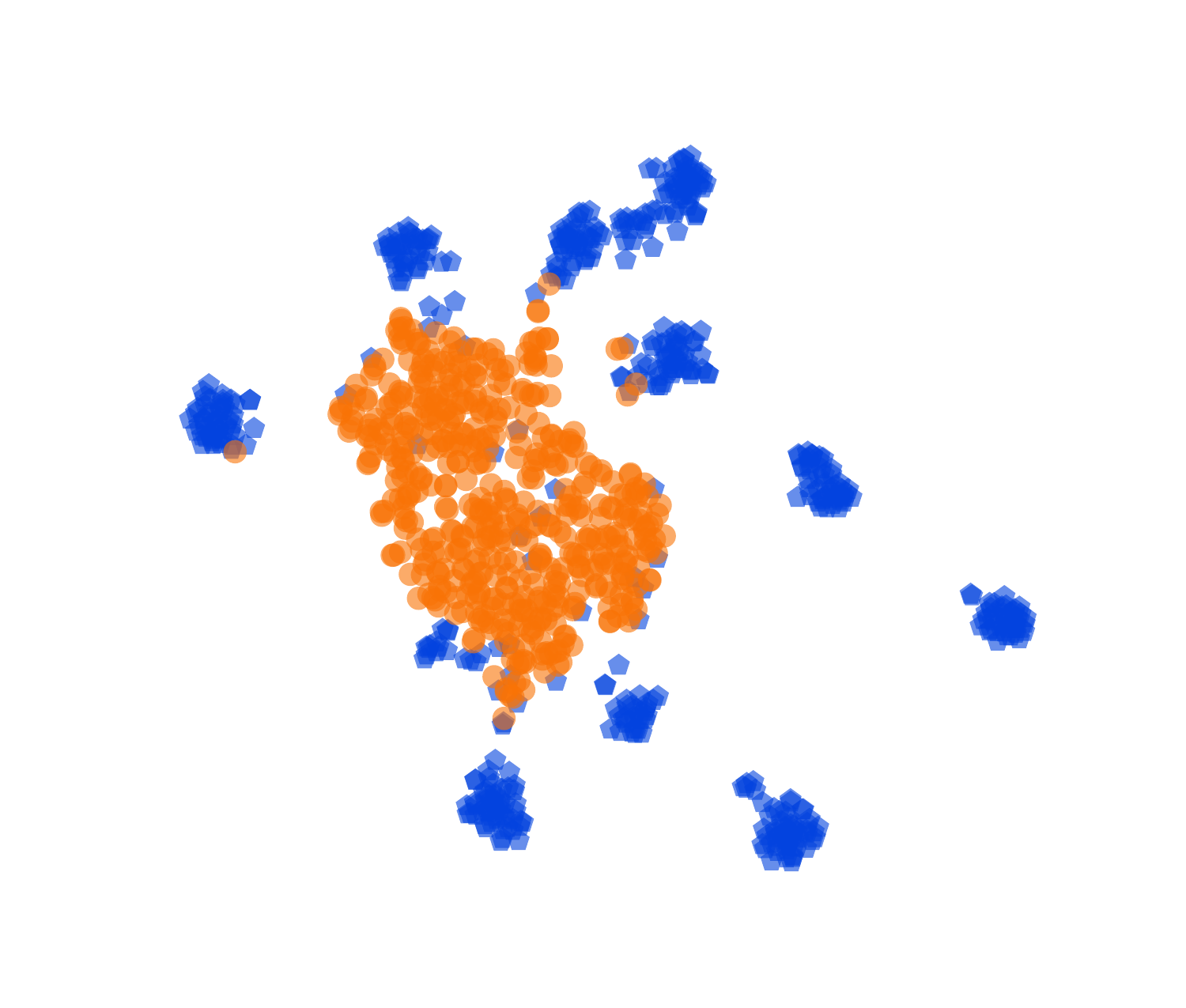}
\caption{DB-DFMS}
\label{figure:db-dfms_tsne_embeddings}
\end{subfigure}
\caption{t-SNE representations for the embedding of 512 randomly generated data samples with different data-free model stealing methods. The Victim model is ResNet-34-8x trained on CIFAR-10 and the clone model is ResNet-18-8x.}
\label{figure:tsne_for_embeddings}
\end{figure*}

We further take the generated data samples as the input to the victim model and visualize the embeddings output from the penultimate layer by using t-Distributed Neighbor Embedding (t-SNE)~\citep{MH08}, which is depicted in \autoref{figure:tsne_for_embeddings}.
The general trend is the better the distribution of generated datasets matches the distribution of the victim datasets, the greater the attack performance is.
Specifically, ``Random Noise'' seems outliers to the dataset used in the victim model, while our attack is able to produce data samples closer to the victim's distribution.
Though these two distributions cannot match each other ideally, the generated distribution is already enough to extract the information for achieving impressive attack performance.

\section{Conclusion}
\label{section:conclusion}
In this paper, we revisit the generator-based data-free model stealing attack from a diversity perspective, and investigate the possibility of simplifying the existing approaches.
We find that the diversity of the generated data samples used for querying the victim model is one of the key points related to the attack performance.
We conduct extensive experiments to show that simply using a diversity loss to train a generator can force the generation of data samples across all the classes and enable the attack to achieve comparable results as the state-of-the-art methods while with much less computational cost.
We further conduct our attack in more realistic scenarios, e.g., the query budget is limited, or the target model architecture is not available. 
Results evince our attack consistently performs well, which demonstrates the practicality of our attack.
Moreover, entropy and visualization of the generated data samples are provided to explain the success of our attack.

We emphasize the role of diversity in generated data samples for the success of model stealing attacks, while also admitting the existence of other factors that will influence the attack performance. 
Thus for future work, one direction is to search for more elegant approaches to further promote the generation of more diverse data, another way will be combining other methods such as gradient-based ones to reach better attack results, or achieve a trade-off between the attack performance and efficiency.

%-------------------------------------------------------------------------------
\bibliographystyle{plain}
\bibliography{main}

\begin{thebibliography}{10}

\bibitem{CIFAR}
\url{https://www.cs.toronto.edu/~kriz/cifar.html}.

\bibitem{ACB17}
Martin Arjovsky, Soumith Chintala, and L{\'e}on Bottou.
\newblock {Wasserstein Generative Adversarial Networks}.
\newblock In {\em {International Conference on Machine Learning (ICML)}}, pages 214--223. PMLR, 2017.

\bibitem{CWXYLSXXT19}
Hanting Chen, Yunhe Wang, Chang Xu, Zhaohui Yang, Chuanjian Liu, Boxin Shi, Chunjing Xu, Chao Xu, and Qi~Tian.
\newblock {Data-Free Learning of Student Networks}.
\newblock In {\em {IEEE International Conference on Computer Vision (ICCV)}}, pages 3513--3521. IEEE, 2019.

\bibitem{CH19}
Jang~Hyun Cho and Bharath Hariharan.
\newblock {On the Efficacy of Knowledge Distillation}.
\newblock In {\em {IEEE International Conference on Computer Vision (ICCV)}}, pages 4793--4801. IEEE, 2019.

\bibitem{CCEL20}
Yoojin Choi, Jihwan~P. Choi, Mostafa El{-}Khamy, and Jungwon Lee.
\newblock {Data-Free Network Quantization With Adversarial Knowledge Distillation}.
\newblock In {\em {IEEE Conference on Computer Vision and Pattern Recognition (CVPR)}}, pages 3047--3057. IEEE, 2020.

\bibitem{GPMXWOCB14}
Ian Goodfellow, Jean Pouget-Abadie, Mehdi Mirza, Bing Xu, David Warde-Farley, Sherjil Ozair, Aaron Courville, and Yoshua Bengio.
\newblock {Generative Adversarial Nets}.
\newblock In {\em {Annual Conference on Neural Information Processing Systems (NIPS)}}, pages 2672--2680. NIPS, 2014.

\bibitem{HZRS16}
Kaiming He, Xiangyu Zhang, Shaoqing Ren, and Jian Sun.
\newblock {Deep Residual Learning for Image Recognition}.
\newblock In {\em {IEEE Conference on Computer Vision and Pattern Recognition (CVPR)}}, pages 770--778. IEEE, 2016.

\bibitem{HVD15}
Geoffrey~E. Hinton, Oriol Vinyals, and Jeffrey Dean.
\newblock {Distilling the Knowledge in a Neural Network}.
\newblock {\em {CoRR abs/1503.02531}}, 2015.

\bibitem{HLMW17}
Gao Huang, Zhuang Liu, Laurens van~der Maaten, and Kilian~Q. Weinberger.
\newblock {Densely Connected Convolutional Networks}.
\newblock In {\em {IEEE Conference on Computer Vision and Pattern Recognition (CVPR)}}, pages 2261--2269. IEEE, 2017.

\bibitem{KPQ21}
Sanjay Kariyappa, Atul Prakash, and Moinuddin~K. Qureshi.
\newblock {{MAZE:} Data-Free Model Stealing Attack Using Zeroth-Order Gradient Estimation}.
\newblock In {\em {IEEE Conference on Computer Vision and Pattern Recognition (CVPR)}}, pages 13814--13823. IEEE, 2021.

\bibitem{KLA19}
Tero Karras, Samuli Laine, and Timo Aila.
\newblock {A Style-Based Generator Architecture for Generative Adversarial Networks}.
\newblock In {\em {IEEE Conference on Computer Vision and Pattern Recognition (CVPR)}}, pages 4401--4410. IEEE, 2019.

\bibitem{LLWT15}
Ziwei Liu, Ping Luo, Xiaogang Wang, and Xiaoou Tang.
\newblock {Deep Learning Face Attributes in the Wild}.
\newblock In {\em {IEEE International Conference on Computer Vision (ICCV)}}, pages 3730--3738. IEEE, 2015.

\bibitem{MS19}
Paul Micaelli and Amos~J. Storkey.
\newblock {Zero-shot Knowledge Transfer via Adversarial Belief Matching}.
\newblock In {\em {Annual Conference on Neural Information Processing Systems (NeurIPS)}}, pages 9547--9557. NeurIPS, 2019.

\bibitem{NMSRC19}
Gaurav~Kumar Nayak, Konda~Reddy Mopuri, Vaisakh Shaj, Venkatesh~Babu Radhakrishnan, and Anirban Chakraborty.
\newblock {Zero-Shot Knowledge Distillation in Deep Networks}.
\newblock In {\em {International Conference on Machine Learning (ICML)}}, pages 4743--4751. PMLR, 2019.

\bibitem{OSF19}
Tribhuvanesh Orekondy, Bernt Schiele, and Mario Fritz.
\newblock {Knockoff Nets: Stealing Functionality of Black-Box Models}.
\newblock In {\em {IEEE Conference on Computer Vision and Pattern Recognition (CVPR)}}, pages 4954--4963. IEEE, 2019.

\bibitem{PMGJCS17}
Nicolas Papernot, Patrick~D. McDaniel, Ian Goodfellow, Somesh Jha, Z.~Berkay Celik, and Ananthram Swami.
\newblock {Practical Black-Box Attacks Against Machine Learning}.
\newblock In {\em {ACM Asia Conference on Computer and Communications Security (ASIACCS)}}, pages 506--519. ACM, 2017.

\bibitem{RPM19}
Nicholas Roberts, Vinay~Uday Prabhu, and Matthew McAteer.
\newblock {Model Weight Theft With Just Noise Inputs: The Curious Case of the Petulant Attacker}.
\newblock {\em {CoRR abs/1912.08987}}, 2019.

\bibitem{SHZZC18}
Mark Sandler, Andrew~G. Howard, Menglong Zhu, Andrey Zhmoginov, and Liang{-}Chieh Chen.
\newblock {MobileNetV2: Inverted Residuals and Linear Bottlenecks}.
\newblock In {\em {IEEE Conference on Computer Vision and Pattern Recognition (CVPR)}}, pages 4510--4520. IEEE, 2018.

\bibitem{SAB22}
Sunandini Sanyal, Sravanti Addepalli, and R.~Venkatesh Babu.
\newblock {Towards Data-Free Model Stealing in a Hard Label Setting}.
\newblock {\em {CoRR abs/2204.11022}}, 2022.

\bibitem{SCDVPB17}
Ramprasaath~R. Selvaraju, Michael Cogswell, Abhishek Das, Ramakrishna Vedantam, Devi Parikh, and Dhruv Batra.
\newblock {Grad-CAM: Visual Explanations from Deep Networks via Gradient-Based Localization}.
\newblock In {\em {IEEE International Conference on Computer Vision (ICCV)}}, pages 618--626. IEEE, 2017.

\bibitem{SSSS17}
Reza Shokri, Marco Stronati, Congzheng Song, and Vitaly Shmatikov.
\newblock {Membership Inference Attacks Against Machine Learning Models}.
\newblock In {\em {IEEE Symposium on Security and Privacy (S\&P)}}, pages 3--18. IEEE, 2017.

\bibitem{SZ15}
Karen Simonyan and Andrew Zisserman.
\newblock {Very Deep Convolutional Networks for Large-Scale Image Recognition}.
\newblock In {\em {International Conference on Learning Representations (ICLR)}}, 2015.

\bibitem{TZJRR16}
Florian Tram{\`e}r, Fan Zhang, Ari Juels, Michael~K. Reiter, and Thomas Ristenpart.
\newblock {Stealing Machine Learning Models via Prediction APIs}.
\newblock In {\em {USENIX Security Symposium (USENIX Security)}}, pages 601--618. USENIX, 2016.

\bibitem{TMWP21}
Jean{-}Baptiste Truong, Pratyush Maini, Robert~J. Walls, and Nicolas Papernot.
\newblock {Data-Free Model Extraction}.
\newblock In {\em {IEEE Conference on Computer Vision and Pattern Recognition (CVPR)}}, pages 4771--4780. IEEE, 2021.

\bibitem{MH08}
Laurens van~der Maaten and Geoffrey Hinton.
\newblock {Visualizing Data using t-SNE}.
\newblock {\em {Journal of Machine Learning Research}}, 2008.

\bibitem{YLZZ19}
Yuanshun Yao, Huiying Li, Haitao Zheng, and Ben~Y. Zhao.
\newblock {Latent Backdoor Attacks on Deep Neural Networks}.
\newblock In {\em {ACM SIGSAC Conference on Computer and Communications Security (CCS)}}, pages 2041--2055. ACM, 2019.

\bibitem{ZK16}
Sergey Zagoruyko and Nikos Komodakis.
\newblock {Wide Residual Networks}.
\newblock In {\em {Proceedings of the British Machine Vision Conference (BMVC)}}. {BMVA} Press, 2016.

\bibitem{ZGMO18}
Han Zhang, Ian~J. Goodfellow, Dimitris~N. Metaxas, and Augustus Odena.
\newblock {Self-Attention Generative Adversarial Networks}.
\newblock {\em {CoRR abs/1805.08318}}, 2018.

\bibitem{ZJPWLS20}
Yuheng Zhang, Ruoxi Jia, Hengzhi Pei, Wenxiao Wang, Bo~Li, and Dawn Song.
\newblock {The Secret Revealer: Generative Model-Inversion Attacks Against Deep Neural Networks}.
\newblock In {\em {IEEE Conference on Computer Vision and Pattern Recognition (CVPR)}}, pages 250--258. IEEE, 2020.

\end{thebibliography}

\clearpage
\appendix
%-------------------------------------------------------------------------------
\section{Training Details}
\label{section:appendix_training}
%-------------------------------------------------------------------------------
\subsection{Datasets}
\mypara{CIFAR-10.}
CIFAR-10 is a benchmark dataset for image classification task. 
It has 10 classes where each class has 5000 and 1000 data samples for training and testing respectively. 
The size of each image is 32$\times$32$\times$3.

\mypara{SVHN.}
SVHN is an image dataset for digits in real scenarios, which has 10 classes for numbers from ``0'' to ``9''. 
There are in total 73257 data samples for training and 26032 for testing. 
It also consists of additional 531131 difficult images as an extra dataset, while in our experiments we don't consider it.

\mypara{CelebA.}
CelebA is a large-scale dataset for face recognition. It contains 202599 number of images and each of them has 40 binary attributes. 
We choose ``Male'', ``Mouth\_Slightly\_Open'' and ``Smiling'' as the target attributes, and it splits the whole dataset into 8 classes and each class at least has 8561 images. 
Thus we randomly select 8000 images from each class to form a balanced dataset, and use 60000 and 4000 among for training and testing respectively. 
We resize each image to 64$\times$64 as they don't have a fixed size.

\subsection{Model Training}
For the victim model, we train the model for 50 epochs on SVHN and 200 epochs on CIFAR-10 and CelebA. 
The optimizer is SGD with an initial learning rate as 0.1, decayed by a cosine scheduler.
The clone model and generator are trained with SGD at 0.1 initial learning rate and Adam at $10^{-4}$ initial learning rate respectively, and both have a batch size of 256 and a scheduler that multiplies the learning rate with 0.3 at 10\%, 30\% and 50\% of the total training epochs.

%-------------------------------------------------------------------------------
\section{Additional Results}
%-------------------------------------------------------------------------------
\subsection{Hyper-parameters Turing}
\mypara{Training Times between Generator and Clone Model.} 
In our experiments, we set $n_\mathcal{G}$ and $n_\mathcal{C}$ to balance the training between the generator and clone model. Here we show the effect of the ratio between these two hyper-parameters on the attack performance in Table \ref{table:different_ng_nc}.

\begin{table}[H]
\centering
\caption{The attack performance of DB-DFMS against ResNet-34-8x trained on CIFAR-10 with different ratio of $n_\mathcal{G}$ and $n_\mathcal{C}$. The accuracy of the victim model is 0.930, and the clone model is ResNet-18-8x.}
\label{table:different_ng_nc}
\setlength{\tabcolsep}{4.0pt}
\scalebox{0.9}{
\begin{tabular}{lcccccccccc}
\toprule
\rowcolor{white}
$n_\mathcal{G}:n_\mathcal{C}$ & 1:1 & 1:2 & 1:3 & 1:4 & 1:5 & 1:6 & 1:7 & 1:8 & 1:9 & 1:10 \\
\midrule
Accuracy & 0.829 & 0.860 & 0.880 & 0.881 & 0.885 & 0.886 & 0.887 & 0.885 & 0.880 & 0.875 \\
Agreement & 0.841 & 0.884 & 0.913 & 0.914 & 0.921 & 0.915 & 0.918 & 0.917 & 0.907 & 0.897 \\ 
\bottomrule
\end{tabular}
}
\end{table}

\mypara{Clone Model Loss Functions.}
As recommended in previous papers, we adopt $l_1$ norm loss for training the clone model. Here we consider different loss functions and see the impact on the attack performance.

\begin{table}[H]
\centering
\caption{The attack performance of DB-DFMS against ResNet-34-8x trained on CIFAR-10 with different loss functions for clone model. The accuracy of the victim model is 0.930, and the clone model is ResNet-18-8x. ``Acc'' and ``Agr'' represent clone model accuracy and agreement between the victim model and clone model respectively.}
\label{table:different_clone_loss}
\setlength{\tabcolsep}{4.0pt}
\scalebox{0.9}{
\begin{tabular}{lcccccc}
\toprule
\rowcolor{white}
Datasets & \multicolumn{2}{c}{KL Divergence} & \multicolumn{2}{c}{$L_2$ Loss} & \multicolumn{2}{c}{$L_1$ Loss}\\
\cmidrule(l{1pt}r{1pt}){2-3}\cmidrule(l{1pt}r{1pt}){4-5}\cmidrule(l{1pt}r{0pt}){6-7}
(budget) & Acc & Agr & Acc & Agr & Acc & Agr\\
\midrule
CIFAR-10 (20M) & 0.758 & 0.783 & 0.851 & 0.880 & 0.885 & 0.921\\
\bottomrule
\end{tabular}
}
\end{table}

\mypara{Batch Size.}
The core idea of our attack is to train a generator by leveraging a diversity loss, and such a diversity loss is calculated as the negative entropy of the predictions from a mini-batch. Thus we consider the effect of batch size as it will influence the information used in the calculated loss.

\begin{table}[H]
\centering
\caption{The attack performance of DB-DFMS against ResNet-34-8x trained on CIFAR-10 with different batch sizes. The accuracy of the victim model is 0.930, and the clone model is ResNet-18-8x.}
\label{table:different_batch_size}
\setlength{\tabcolsep}{4.0pt}
\scalebox{0.9}{
\begin{tabular}{lcccccccccc}
\toprule
\rowcolor{white}
Batch Size & 16 & 32 & 64 & 128 & 200 & 256 & 300 & 400 & 512 & 1024 \\
\midrule
Accuracy & 0.790 & 0.849 & 0.874 & 0.875 & 0.890 & 0.885 & 0.887 & 0.867 & 0.860 & 0.829\\
Agreement & 0.815 & 0.876 & 0.904 & 0.909 & 0.928 & 0.921 & 0.923 & 0.895 & 0.888 & 0.849\\
\bottomrule
\end{tabular}
}
\end{table}

\subsection{Visualization Results}

\begin{minipage}{0.45\textwidth}
\centering
\captionsetup{type=figure}
\captionsetup[subfigure]{labelformat=empty}
\hspace{-1mm}
\subfloat{\includegraphics[width=0.11\columnwidth]{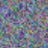}
\includegraphics[width=0.11\columnwidth]{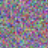}
\includegraphics[width=0.11\columnwidth]{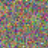}
\includegraphics[width=0.11\columnwidth]{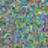}}
\hspace{1mm}
\subfloat{\includegraphics[width=0.11\columnwidth]{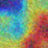}
\includegraphics[width=0.11\columnwidth]{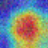}
\includegraphics[width=0.11\columnwidth]{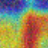}
\includegraphics[width=0.11\columnwidth]{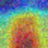}}

\hspace{-1mm}
\subfloat{\includegraphics[width=0.11\columnwidth]{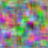}
\includegraphics[width=0.11\columnwidth]{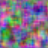}
\includegraphics[width=0.11\columnwidth]{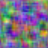}
\includegraphics[width=0.11\columnwidth]{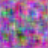}}
\hspace{1mm}
\subfloat{\includegraphics[width=0.11\columnwidth]{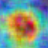}
\includegraphics[width=0.11\columnwidth]{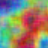}
\includegraphics[width=0.11\columnwidth]{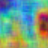}
\includegraphics[width=0.11\columnwidth]{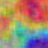}}

\hspace{-1mm}
\subfloat{\includegraphics[width=0.11\columnwidth]{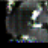}
\includegraphics[width=0.11\columnwidth]{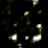}
\includegraphics[width=0.11\columnwidth]{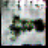}
\includegraphics[width=0.11\columnwidth]{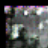}}
\hspace{1mm}
\subfloat{\includegraphics[width=0.11\columnwidth]{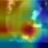}
\includegraphics[width=0.11\columnwidth]{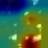}
\includegraphics[width=0.11\columnwidth]{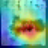}
\includegraphics[width=0.11\columnwidth]{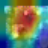}}

\hspace{-1mm}
\subfloat[Generated Images]{\includegraphics[width=0.11\columnwidth]{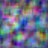}
\includegraphics[width=0.11\columnwidth]{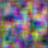}
\includegraphics[width=0.11\columnwidth]{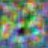}
\includegraphics[width=0.11\columnwidth]{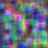}}
\hspace{1mm}
\subfloat[Grad Cam]{\includegraphics[width=0.11\columnwidth]{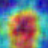}
\includegraphics[width=0.11\columnwidth]{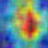}
\includegraphics[width=0.11\columnwidth]{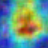}
\includegraphics[width=0.11\columnwidth]{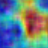}}

\vspace{16pt}
\caption{Generated data Samples and visualization of Grad-CAM from Random Noise, DFME, DFMS-SL and DB-DFMS (from top to bottom) For models trained on SVHN.}
\label{figure:svhn_generated_images}
\end{minipage}
\hfill
\begin{minipage}{0.45\textwidth}
\centering
\captionsetup{type=figure}
\captionsetup[subfigure]{labelformat=empty}
\hspace{-1mm}
\subfloat{\includegraphics[width=0.11\columnwidth]{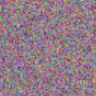}
\includegraphics[width=0.11\columnwidth]{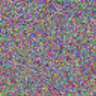}
\includegraphics[width=0.11\columnwidth]{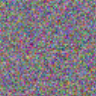}
\includegraphics[width=0.11\columnwidth]{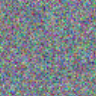}}
\hspace{1mm}
\subfloat{\includegraphics[width=0.11\columnwidth]{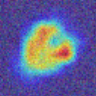}
\includegraphics[width=0.11\columnwidth]{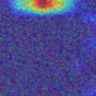}
\includegraphics[width=0.11\columnwidth]{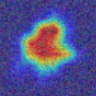}
\includegraphics[width=0.11\columnwidth]{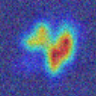}}

\hspace{-1mm}
\subfloat{\includegraphics[width=0.11\columnwidth]{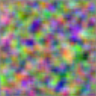}
\includegraphics[width=0.11\columnwidth]{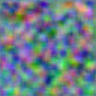}
\includegraphics[width=0.11\columnwidth]{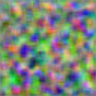}
\includegraphics[width=0.11\columnwidth]{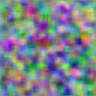}}
\hspace{1mm}
\subfloat{\includegraphics[width=0.11\columnwidth]{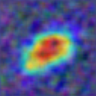}
\includegraphics[width=0.11\columnwidth]{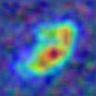}
\includegraphics[width=0.11\columnwidth]{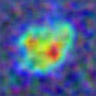}
\includegraphics[width=0.11\columnwidth]{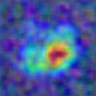}}

\hspace{-1mm}
\subfloat{\includegraphics[width=0.11\columnwidth]{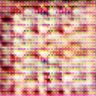}
\includegraphics[width=0.11\columnwidth]{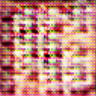}
\includegraphics[width=0.11\columnwidth]{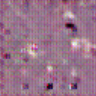}
\includegraphics[width=0.11\columnwidth]{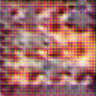}}
\hspace{1mm}
\subfloat{\includegraphics[width=0.11\columnwidth]{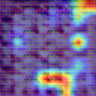}
\includegraphics[width=0.11\columnwidth]{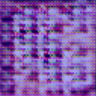}
\includegraphics[width=0.11\columnwidth]{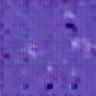}
\includegraphics[width=0.11\columnwidth]{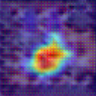}}

\hspace{-1mm}
\subfloat[Generated Images]{\includegraphics[width=0.11\columnwidth]{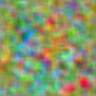}
\includegraphics[width=0.11\columnwidth]{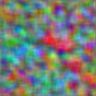}
\includegraphics[width=0.11\columnwidth]{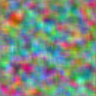}
\includegraphics[width=0.11\columnwidth]{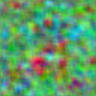}}
\hspace{1mm}
\subfloat[Grad Cam]{\includegraphics[width=0.11\columnwidth]{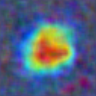}
\includegraphics[width=0.11\columnwidth]{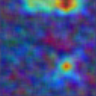}
\includegraphics[width=0.11\columnwidth]{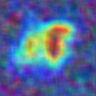}
\includegraphics[width=0.11\columnwidth]{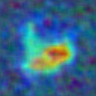}}

\vspace{16pt}
\caption{Generated data Samples and visualization of Grad-CAM from Random Noise, DFME, DFMS-SL and DB-DFMS (from top to bottom) For models trained on CelebA.}
\label{figure:CelebA_generated_images}
\end{minipage}

\begin{figure*}[!ht]
\centering
\begin{subfigure}{0.245\columnwidth}
\includegraphics[width=\columnwidth]{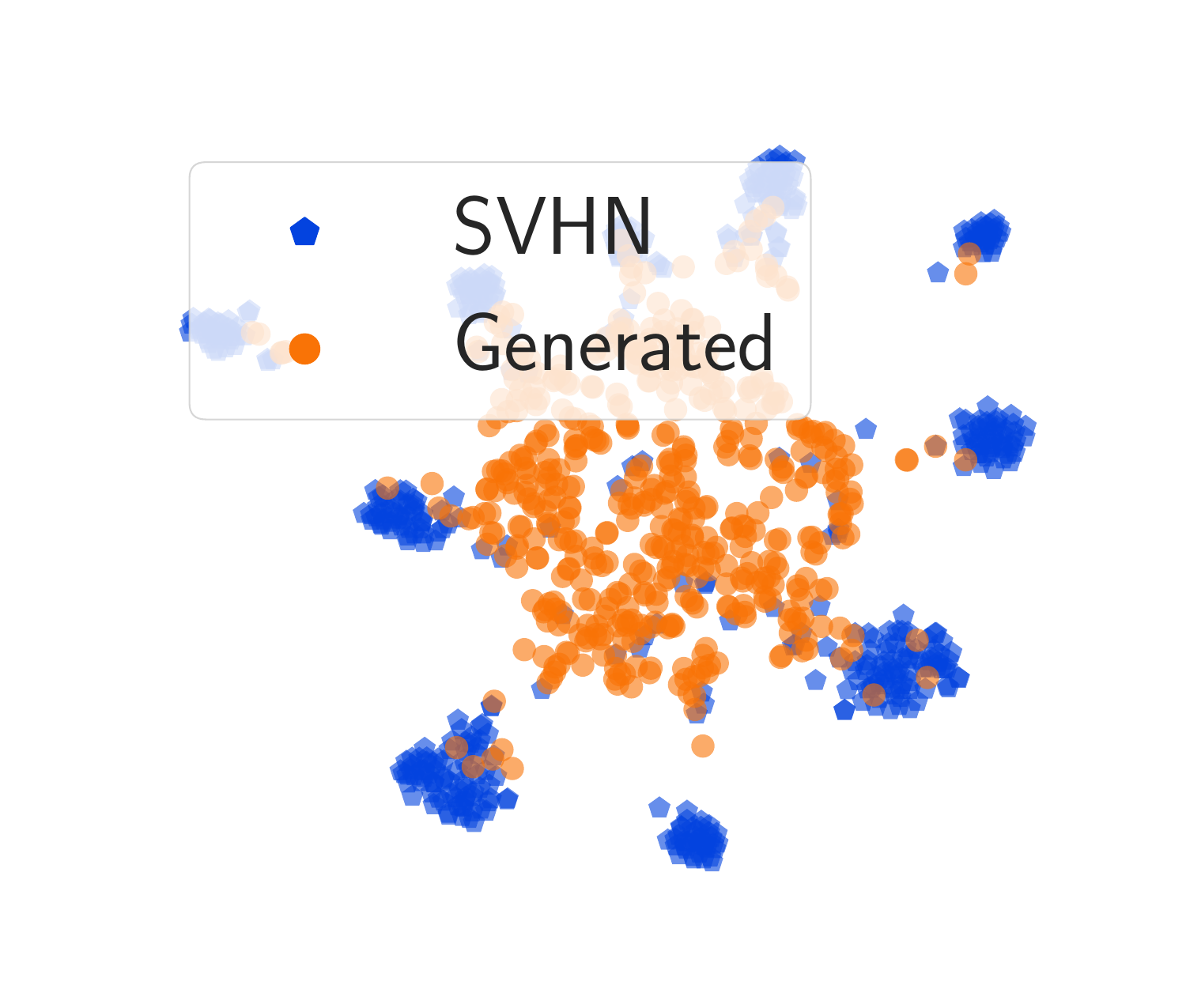}
\caption{Random Noise}
\label{figure:noise_tsne_embeddings_svhn}
\end{subfigure}
\begin{subfigure}{0.245\columnwidth}
\includegraphics[width=\columnwidth]{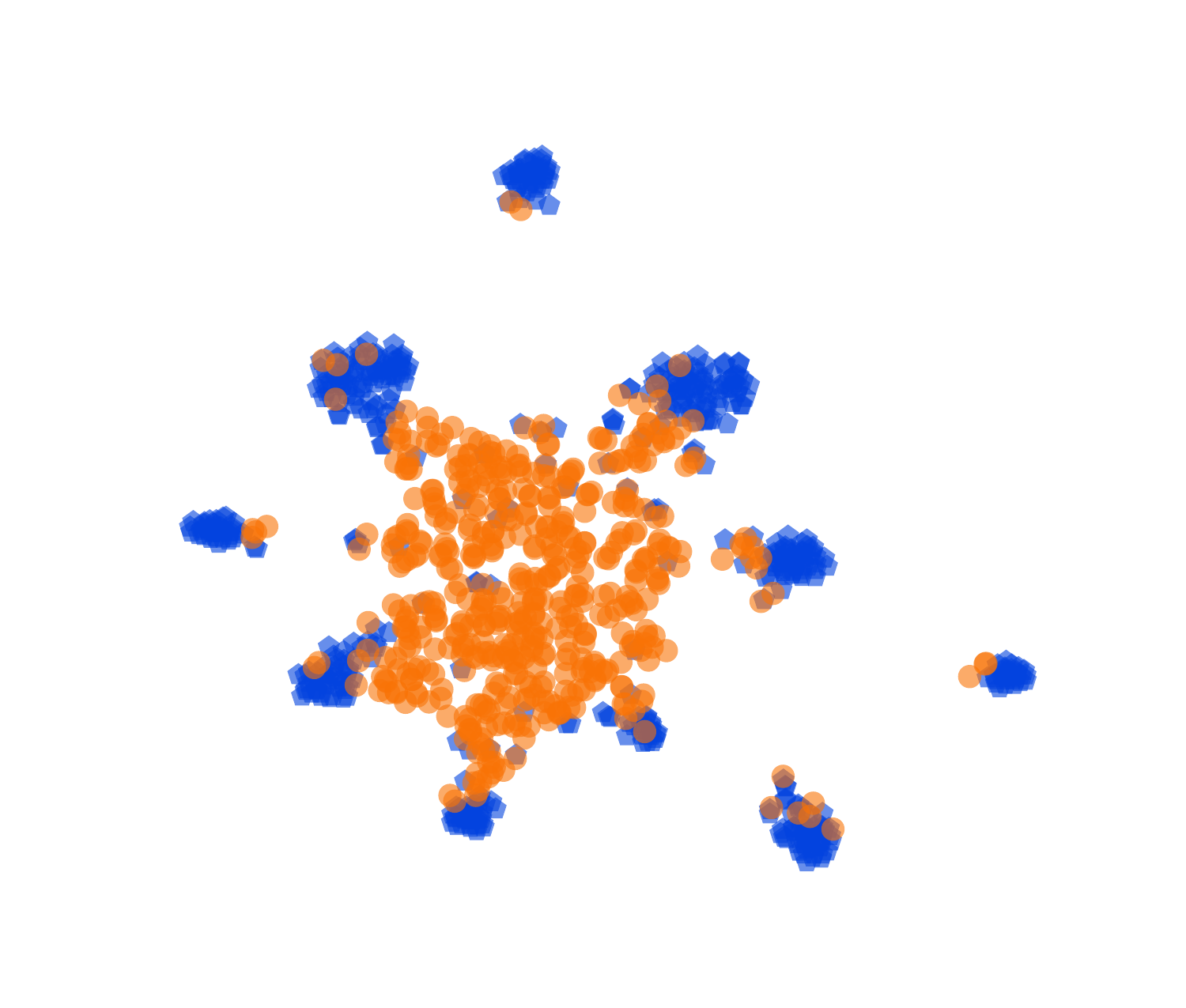}
\caption{DFME}
\label{figure:dfme_tsne_embeddings_svhn}
\end{subfigure}
\begin{subfigure}{0.245\columnwidth}
\includegraphics[width=\columnwidth]{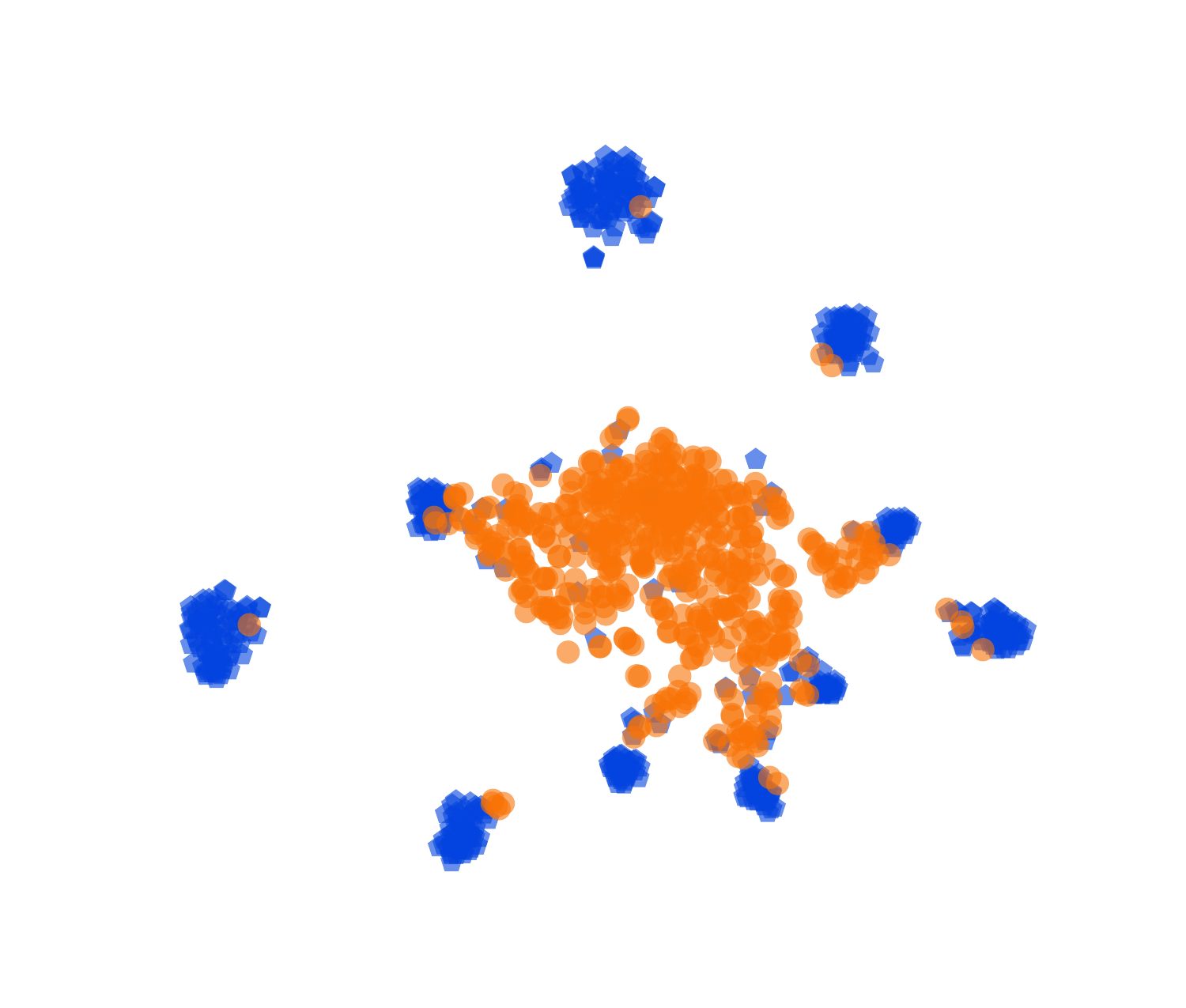}
\caption{DFMS-SL}
\label{figure:dfms_tsne_embeddings_svhn}
\end{subfigure}
\begin{subfigure}{0.245\columnwidth}
\includegraphics[width=\columnwidth]{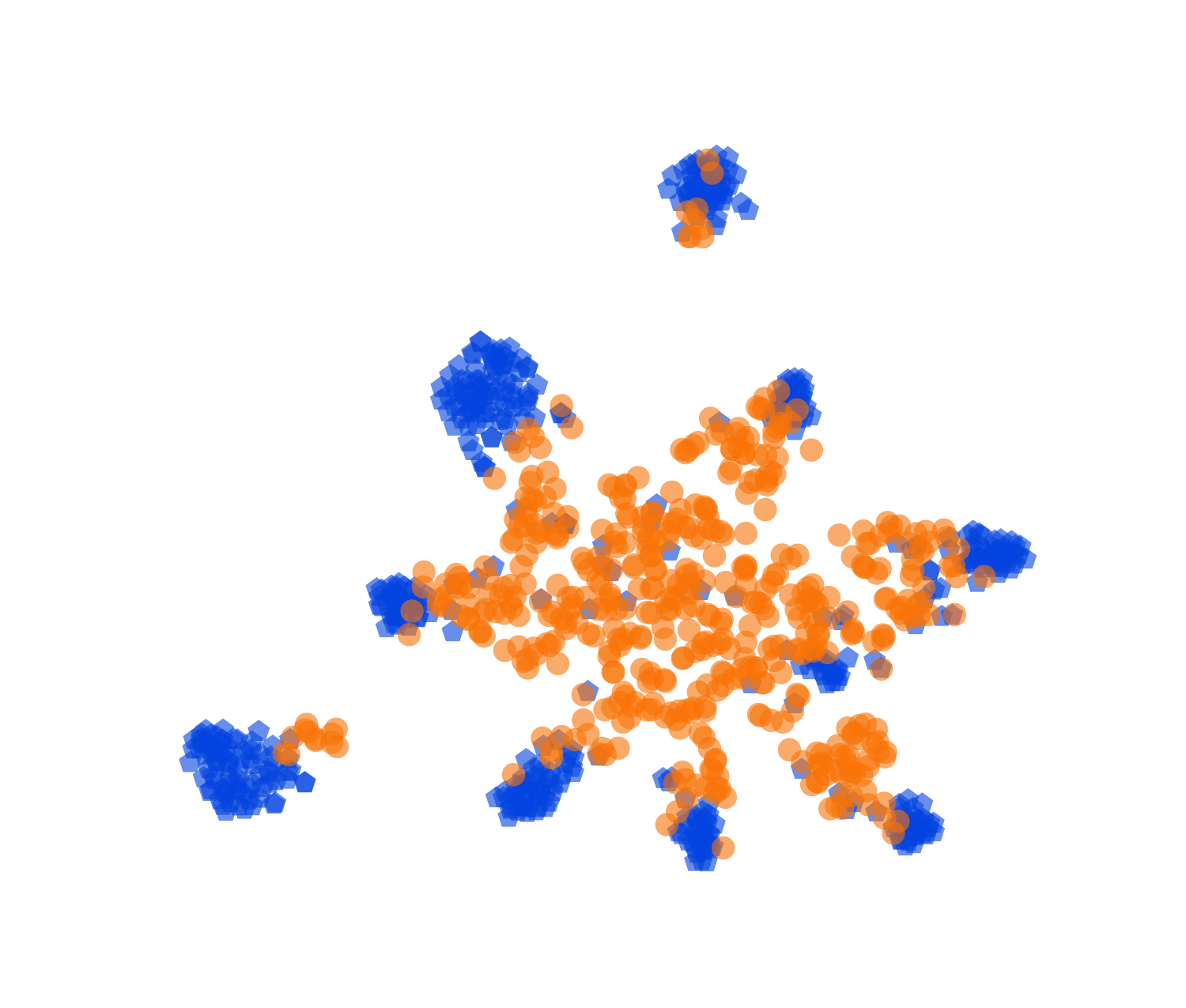}
\caption{DB-DFMS}
\label{figure:db-dfms_tsne_embeddings_svhn}
\end{subfigure}
\caption{t-SNE representations for the embedding of 512 randomly generated data samples with different data-free model stealing methods. The Victim model is ResNet-34-8x trained on SVHN and the clone model is ResNet-18-8x.}
\label{figure:tsne_for_embeddings_svhn}
\end{figure*}

\begin{figure*}[!ht]
\centering
\begin{subfigure}{0.245\columnwidth}
\includegraphics[width=\columnwidth]{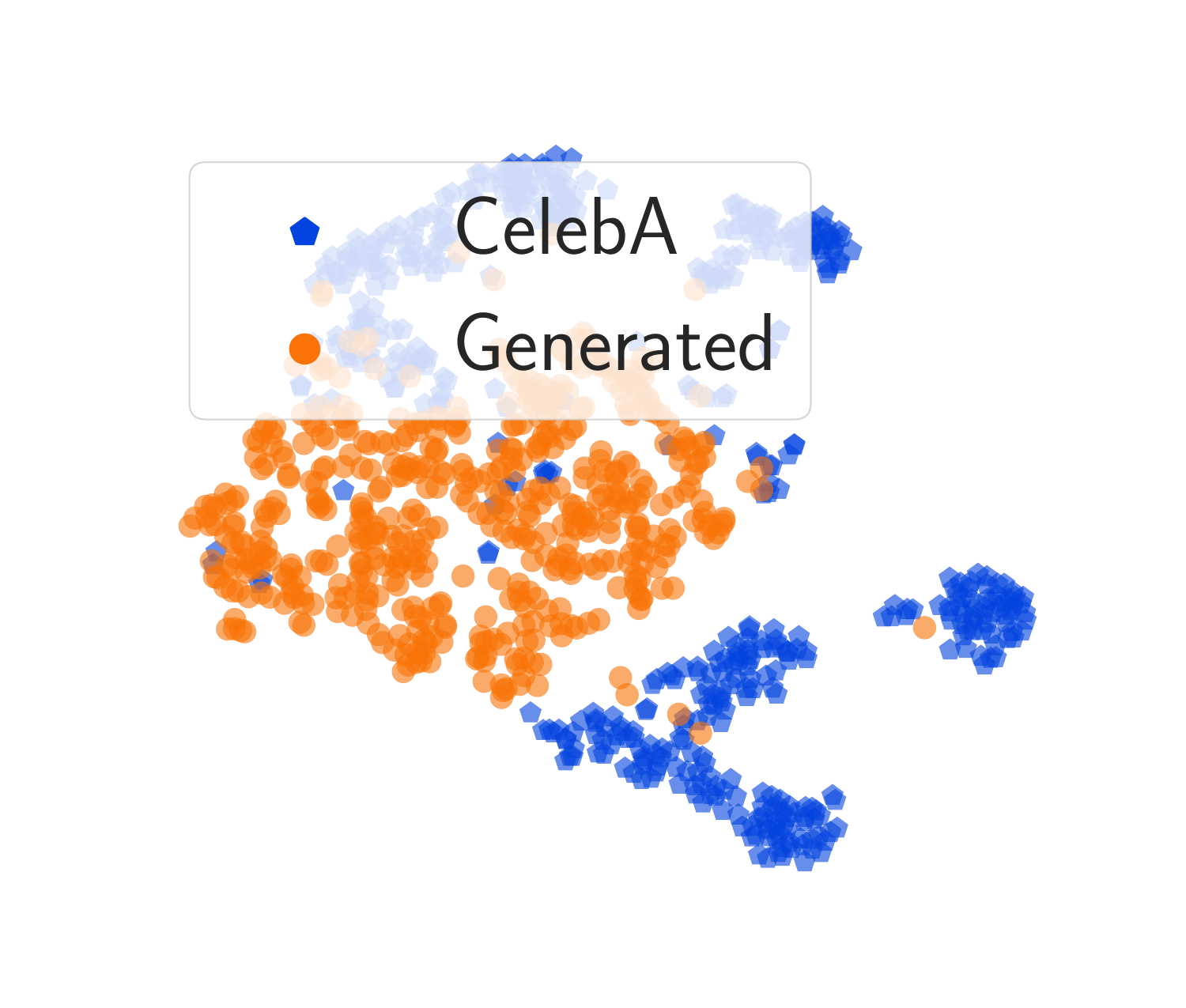}
\caption{Random Noise}
\label{figure:noise_tsne_embeddings_celeba}
\end{subfigure}
\begin{subfigure}{0.245\columnwidth}
\includegraphics[width=\columnwidth]{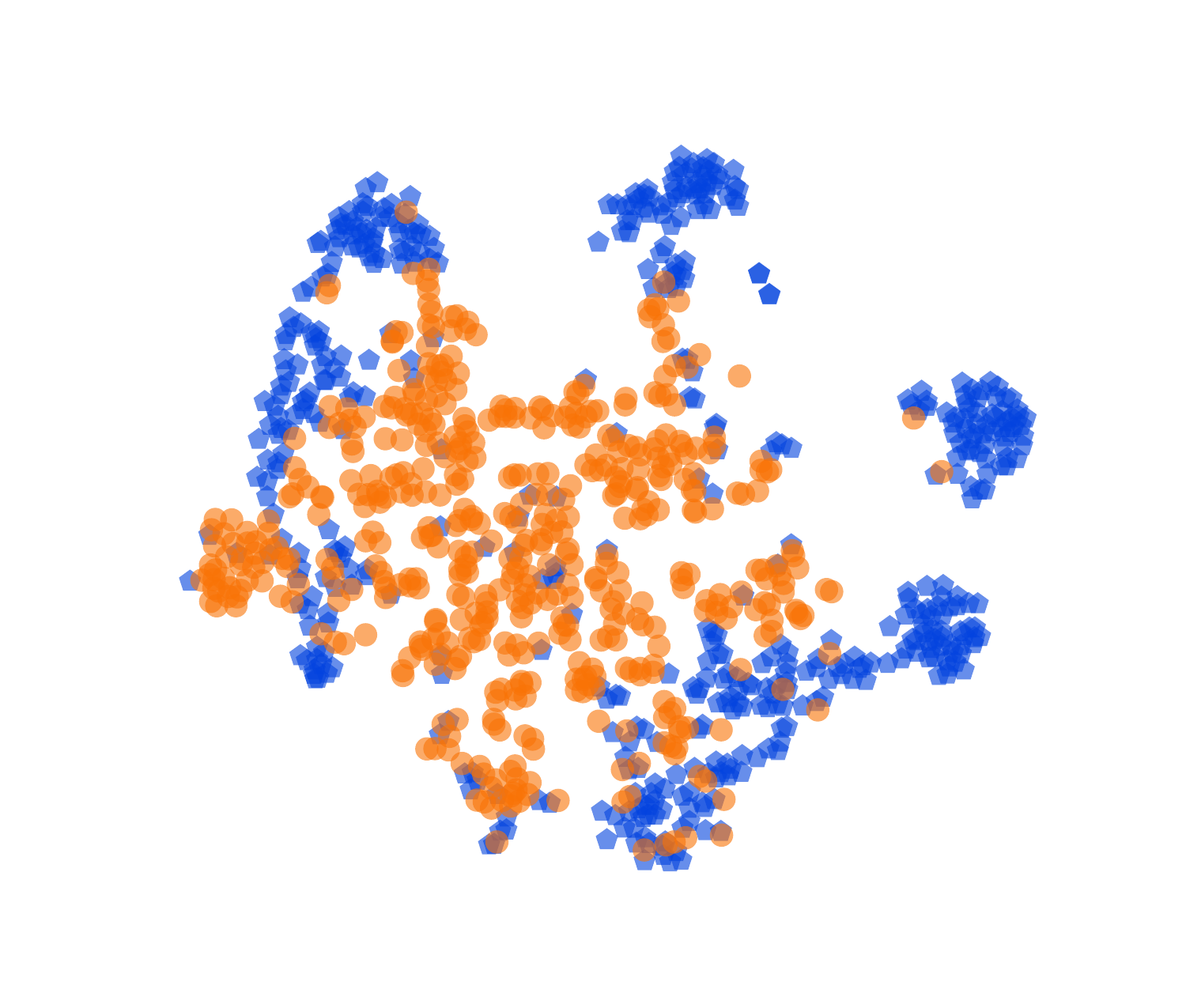}
\caption{DFME}
\label{figure:dfme_tsne_embeddings_celeba}
\end{subfigure}
\begin{subfigure}{0.245\columnwidth}
\includegraphics[width=\columnwidth]{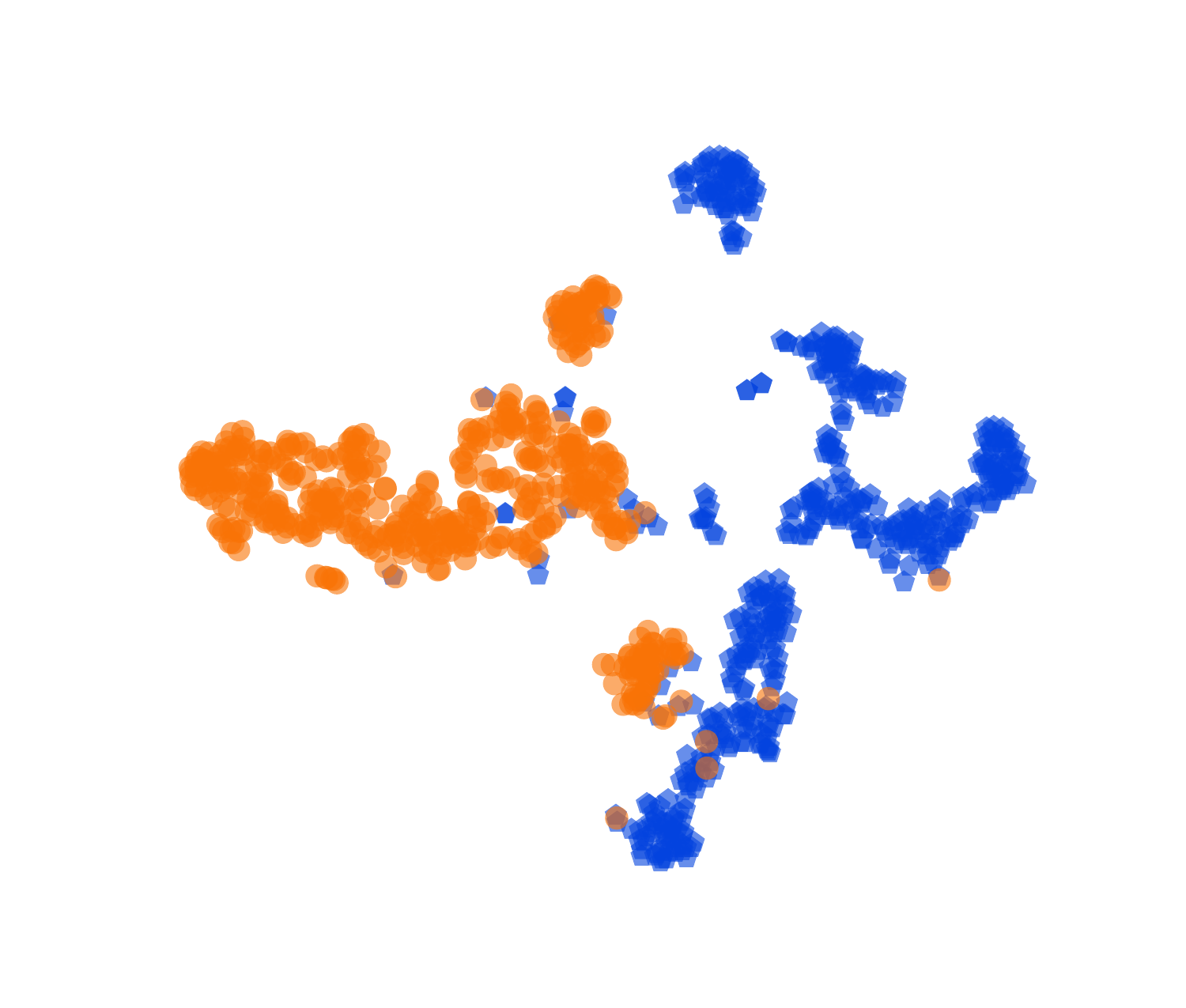}
\caption{DFMS-SL}
\label{figure:dfms_tsne_embeddings_celeba}
\end{subfigure}
\begin{subfigure}{0.245\columnwidth}
\includegraphics[width=\columnwidth]{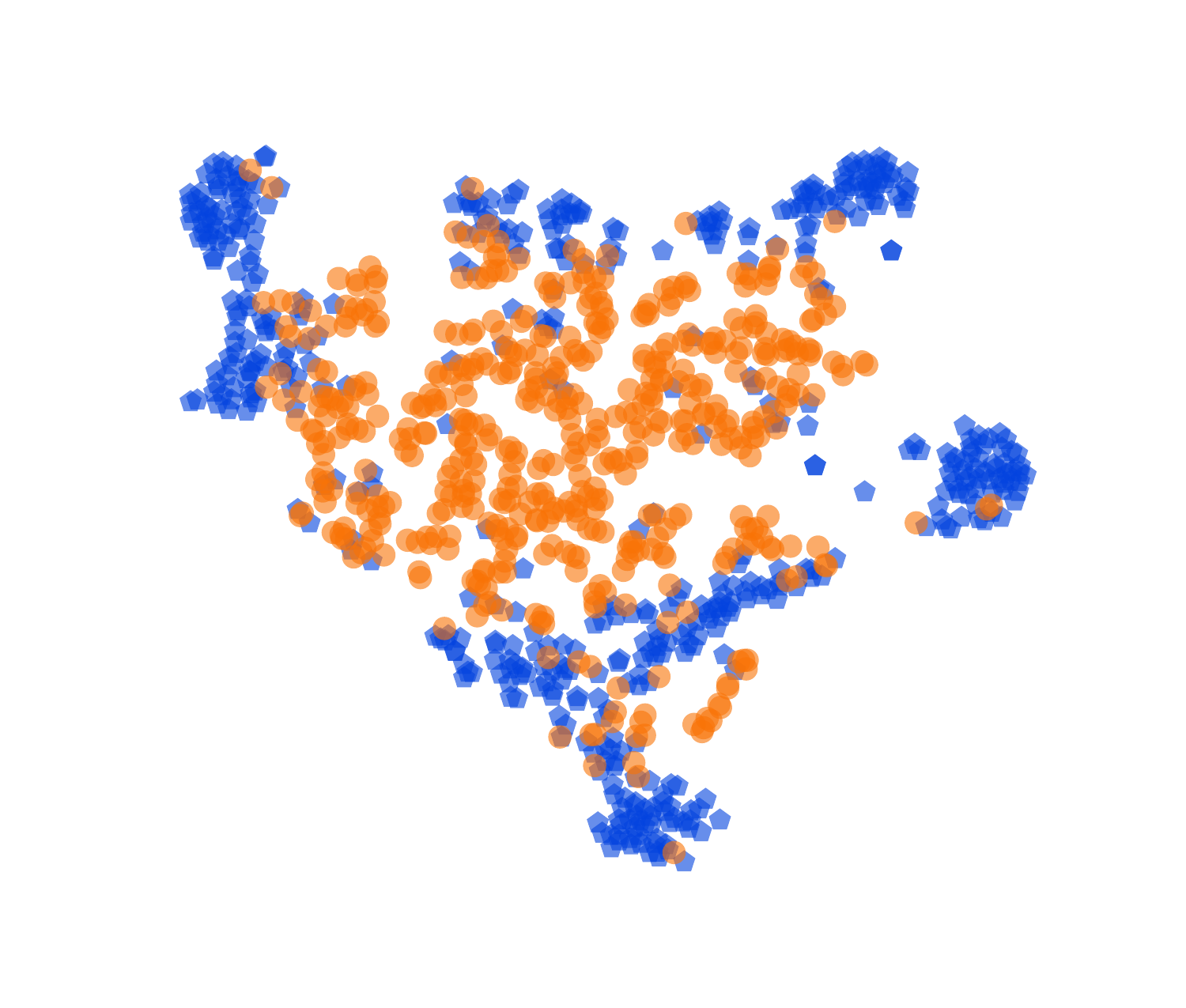}
\caption{DB-DFMS}
\label{figure:db-dfms_tsne_embeddings_celeba}
\end{subfigure}
\caption{t-SNE representations for the embedding of 512 randomly generated data samples with different data-free model stealing methods. The Victim model is ResNet-34-8x trained on CelebA and the clone model is ResNet-18-8x.}
\label{figure:tsne_for_embeddings_celeba}
\end{figure*}

Here we provide the visualization for SVHN and CelebA, and the results follow the patterns we claim in the main experiments. For SVHN, all methods perform well as it is a simple task, thus the embedding of the generated data samples is more aligned to what from the victim model training data, including ``Random Noise'', as shown in Figure \ref{figure:tsne_for_embeddings_svhn}.

\end{document}